\documentclass[manuscript,NETN]{stjour}
\articletype{Research}

\usepackage{amsmath,amssymb}

\usepackage{MnSymbol}
\usepackage{color}
\newcommand{\vdashv}{\vdash\mathrel{\mkern-8mu}\dashv}

\begin{document}
\title{Potential role of a ventral nerve cord central pattern generator in forward and backward locomotion in \textit{Caenorhabditis elegans}}
% \title[A ventral nerve cord CPG could underlie locomotion in \textit{C. elegans}]{A ventral nerve cord CPG could underlie forward and backward locomotion in \LARGE{\emph{Caenorhabditis elegans}}}
% \subtitle{<Subtitle Here>} %% Optional subtitle

%% If shortened title for running head is needed so that the article title can fit
%%   in the running head, use [] argument, ie,
%%
%%   \title[Shortened Title for Running Head]{Title of Article}
%%   \subtitle{Subtitle Here}

%%% Author/Affil
%% Since we use \affil{} in the argument of the \author command, we
%% need to supply another version of the author names, without \affil{}
%% to be used for running heads:

\author[Author Names]% shortened version for running head, optional
{Erick O. Olivares\affil{1},
Eduardo J. Izquierdo\affil{1,2},
Randall D. Beer\affil{1,2}}

\affiliation{1}{Cognitive Science Program, Indiana University, Bloomington, IN. USA}
\affiliation{2}{School of Informatics and Computing, Bloomington, IN. USA}

%ie.
%\affiliation{1}{Gatsby Computational Neuroscience Unit, University
%College London, London, United Kingdom} 

\correspondingauthor{Eduardo J. Izquierdo}{edizquie@indiana.edu}

% ie,
%\correspondingauthor{Ritwik K. Niyogi}{ritwik.niyogi@gatsby.ucl.ac.uk}


%ie
%\keywords{work, leisure, normative, microscopic,  reinforcement learning, economics}

\begin{abstract}
\textit{C.~elegans} locomotes in an undulatory fashion, generating thrust by propagating dorsoventral bends along its body. While central pattern generators (CPGs) are typically involved in animal locomotion, their presence in \textit{C.~elegans} has been questioned, mainly because there has been no evident circuit that supports intrinsic network oscillations. With a fully reconstructed connectome, the question of whether it is possible to have a CPG in the ventral nerve cord (VNC) of \textit{C.~elegans} can be answered through computational models. We modeled a repeating neural unit based on segmentation analysis of the connectome. We then used an evolutionary algorithm to determine the unknown physiological parameters of each neuron so as to match the features of the neural traces of the worm during forward and backward locomotion. We performed 1000 evolutionary runs and consistently found configurations of the neural circuit that produced oscillations matching the main characteristic observed in experimental recordings. In addition to providing an existence proof for the possibility of a CPG in the VNC, we suggest a series of testable hypotheses about its operation. More generally, we show the feasibility and fruitfulness of a methodology to study behavior based on a connectome, in the absence of complete  neurophysiological details.
\end{abstract}

\begin{authorsummary}
Despite the relative simplicity of \textit{C.~elegans}, its locomotion machinery is not yet well understood.  We focus on the generation of dorsoventral body bends. While network central pattern generators are commonly involved in animal locomotion, their presence in \textit{C.~elegans} has been questioned due to a lack of an evident neural circuit to support it.  We developed a computational model grounded in the available neuroanatomy and neurophysiology and we used an evolutionary algorithm to explore the space of possible configurations of the circuit that matched the neural traces observed during forward and backward locomotion in the worm. Our results demonstrate that it is possible for the rhythmic contraction to be produced by a circuit present in the ventral nerve cord.
\end{authorsummary}

% <body of article>
%%%%%%%%%%%%%%%%% 
%%%%%%%%%%%%%%%%%%%%%%%%%%%%%%%%%%%%%%%%%%%%%%%%%%%%%%%%%%%%%%%
% \linenumbers

\newpage
% Use "Eq" instead of "Equation" for equation citations.
\section*{Introduction}

With 302 neurons and an almost completely reconstructed connectome~\citep{White1986,Chen2006,Varshney2011,Jarrell2012}, \textit{Caernohabditis elegans} is an ideal organism to study the relationship between network structure, network dynamics, and behavior. Since nearly the entire behavioral repertoire of \textit{C. elegans} is ultimately expressed through movement, understanding the neuromechanical basis of locomotion and its modulation is especially critical as a foundation upon which analyses of all other behaviors must build. Similar to other nematodes, \textit{C. elegans} locomotes in an undulatory fashion, generating thrust by propagating dorso-ventral bends along its body with a wavelength of undulation that depends on the properties of the medium through which it moves~\citep{Gray1964,Berri2009,Fang2010,Lebois2012}. Movement is generated by a total of 95 rhomboid body wall muscles arranged in staggered pairs along four dorsal/ventral and left/right longitudinal bundles (DR, DL, VR and VL containing 24, 24, 24 and 23 muscles, respectively)~\citep{Sulston1977,Waterston1988}. The neck and body are driven by 75 ventral cord motor neurons divided into 8 distinct classes (dorsal 11-AS, 9-DA, 7-DB, and 6-DD and ventral 12-VA, 11-VB, 6-VC, and 13-VD). The ventral cord is interconnected by a network of chemical and electrical synapses, which are in turn activated by a set of five lateral pairs of command interneurons (AVA, AVB, AVD, AVE, and PVC) that interact with a variety of interneuronal and sensory neurons in the head~\citep{White1976,White1986}. The motor pattern can be divided into two components that need to be understood: the generation of rhythmic alternating bends and the propagation of these bends along the body. Three hypotheses have been proposed for the generation of the rhythmic pattern~\citep{Gjorgjieva2014,Cohen2014,Zhen2015}: (a) Oscillations through stretch-receptor feedback; (b) Oscillations originating from a single CPG in the head circuit; and (c) Oscillations from one or more CPGs along the ventral nerve cord.  To date there have been models of locomotion that range from purely neural~\citep{Bryden2008,Karbowski2008,Niebur1993} to purely mechanical~\citep{Majmudar2012,Niebur1991}. They can be broadly divided into those that include a neuronal central pattern generator (CPG) in the head~\citep{Karbowski2008,Niebur1991,Sakata2005,Wen2012,Deng2016} and those that rely primarily on sensory feedback mechanisms to generate the rhythmic pattern~\citep{Boyle2008,Boyle2012,Bryden2008,Kunert2017}. Until now, no neuroanatomically-grounded model had considered the third hypothesis: the existence of CPGs in the ventral nerve cord.

While CPGs are involved in animal locomotion in many organisms, from leech to humans~\citep{Dimitrijevic1998,Marder2005,Ijspeert2008, Selverston2010,Guertin2013}, their presence in \textit{C.~elegans} has been questioned on two accounts.  First, there has been evidence for the role of stretch receptors in propagating the oscillatory wave posteriorly along the body~\citep{Wen2012}.  However, the VNC CPGs hypothesis cannot be discarded based purely on the participation of stretch receptors. Even with their involvement, intrinsic network oscillations remain a possibility in the worm. 
Second, there has been no clear evidence of pacemaker neurons involved in the generation of oscillations for locomotion in \textit{C.~elegans}. However, regardless of the presence or absence of pacemaker neurons, the possibility of a network oscillator is still feasible. However, work in the locomotion literature has claimed that the synaptic connectivity of the VNC does not contain evident subcircuits that could be interpreted as local CPG elements capable of spontaneously generating oscillatory activity ~\citep{Wen2012}. With a fully reconstructed connectome, the question of whether such subcircuits exist can be explored more systematically through computational models.
 
\begin{figure}[!ht]
\includegraphics[width=8.7cm]{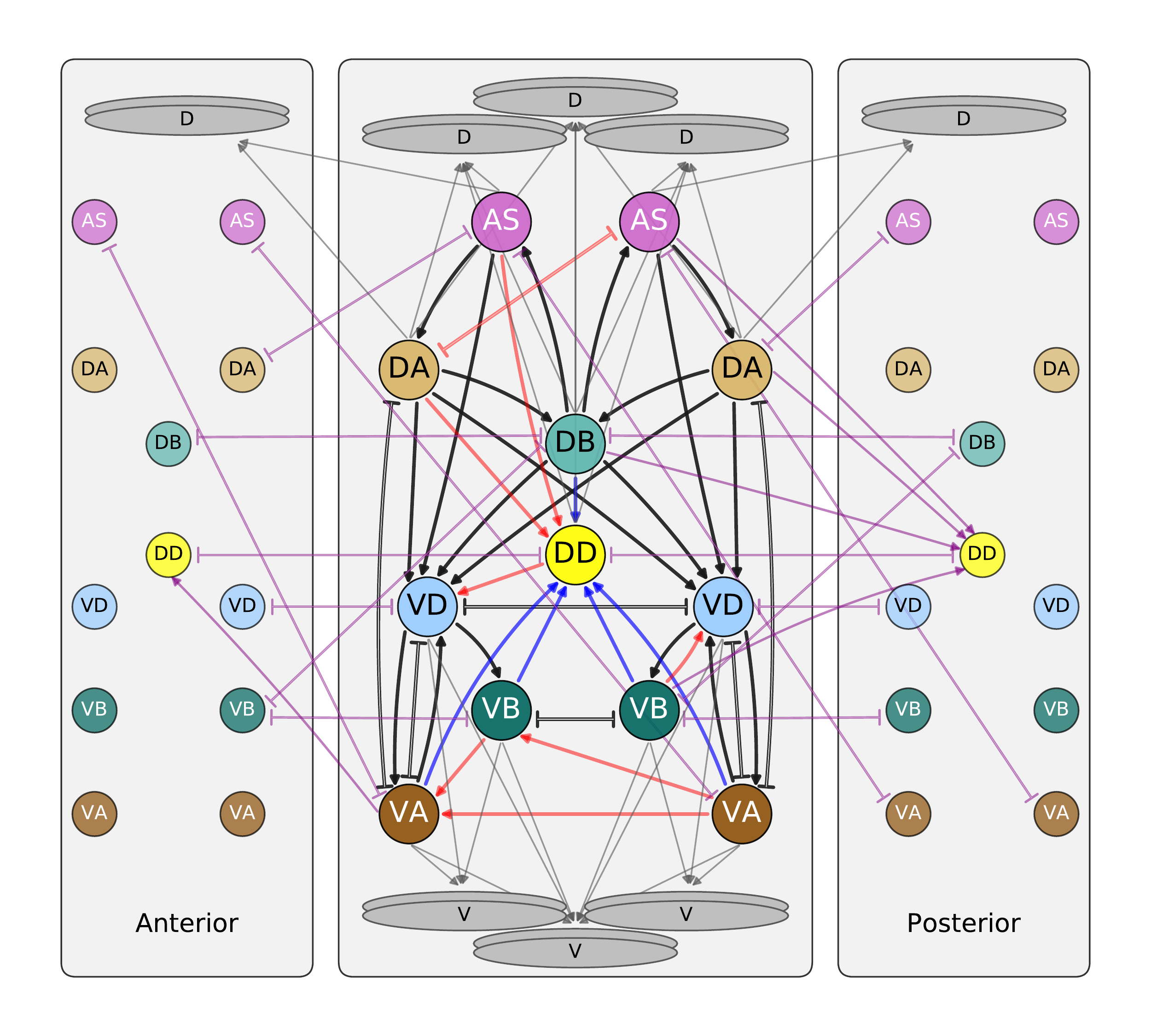}
\centering
\caption{{\bf Repeating neural unit in the VNC, adapted from~\citep{Haspel2011}.}
Arrows represent chemical synapses ($\to$). Connections ending in lines represent electrical synapses ($\vdashv$). Black and blue connections are anterior-posterior symmetric; red connections are not. Purple connections correspond to inter-segmental synapses. Gray connections represents neuromuscular junctions.}
\label{fig_HaspelSegment}
\end{figure}

In order to explore the feasibility of VNC CPGs for locomotion, we developed a computational model of a neural circuit representing one of several repeating neural units present in the VNC. Although the nematode is not segmented, a statistical analysis by Haspel \& O'Donovan of the motorneurons in relation to the position of the muscles they innervate revealed a repeating neural unit along the VNC~\citep{Haspel2011}. Their statistical analysis resulted in a repeating neural unit comprising 2 DA, 1 DB, 2 AS, 2 VD, 2VA, and 1 DD motorneurons, connected by a set of chemical synapses ($\to$) and gap junctions ($\vdashv$) (Fig~\ref{fig_HaspelSegment}). Although not all the connections in this statistically-repeating neural unit are present in every one of the units of the VNC, we used it as a starting point for our study of intrinsic network oscillations. Following previous work~\citep{Izquierdo2010,Izquierdo2013}, motorneurons were modeled as isopotential nodes with the ability to produce a regenerative response, parametrized to match the full range of electrophysiological activity observed in \textit{C.~elegans} neurons~\citep{Mellem2008}.

%%%%%%%%%%%%  Figure Segment as in H&O'D 2011     %%%%%%%%%%%%

\begin{figure}[!ht]
\includegraphics[width=13.2cm]{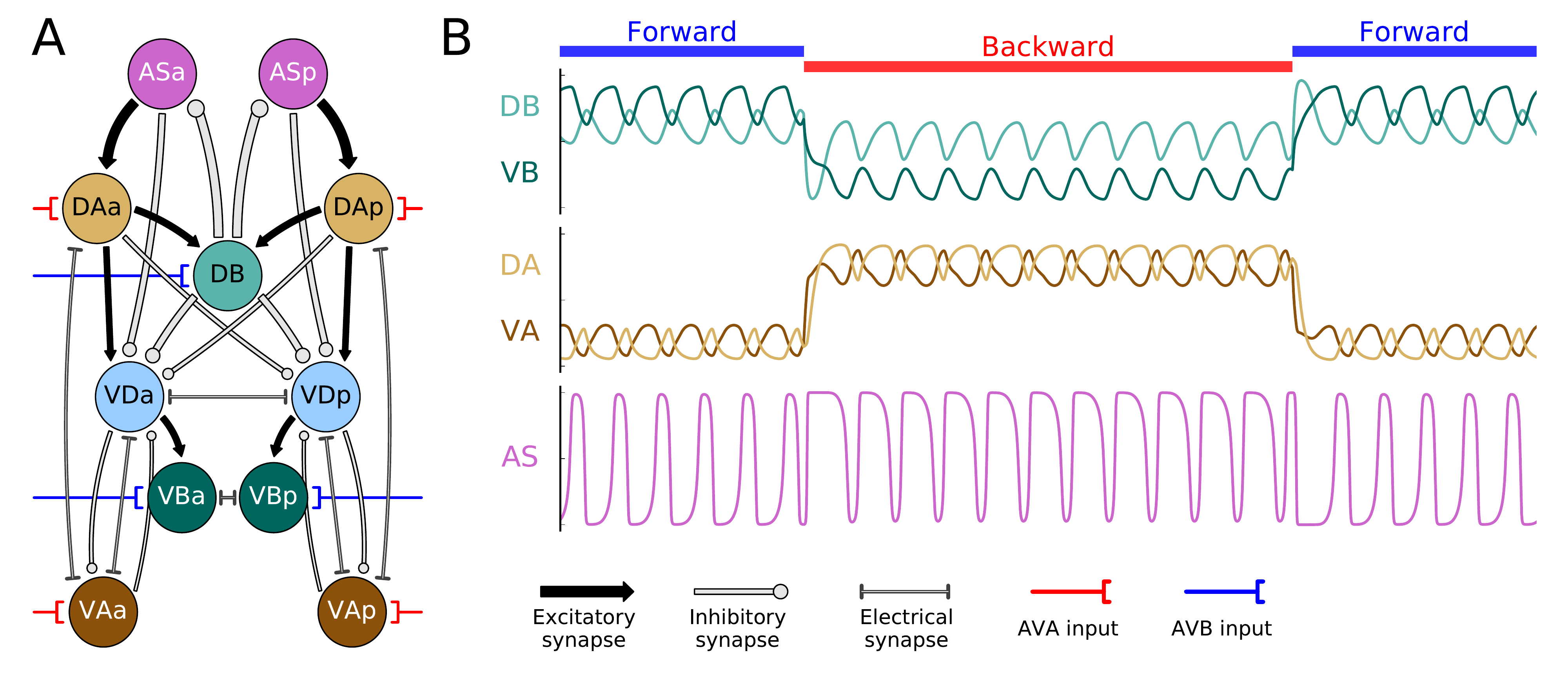}
\centering
\caption{{\bf Best evolved circuit.}
[A]~Full circuit architecture. Neurons and connections taken from VNC neural unit (see Fig.~\ref{fig_HaspelSegment}). The type of connection is depicted with different symbols (see legend). Unknown parameters of the circuit such as the strength of the chemical synapses and sign (whether they are excitatory or inhibitory), the strength of the gap junctions, and the intrinsic parameters of the neurons were evolved. The width of the connection is proportional to the evolved strength. External input from command interneurons AVB to B-class motoneurons and AVA to A-class motoneurons are represented in blue and red, respectively. [B]~Simulated neural traces of the evolved network for forward and backward locomotion. During forward locomotion, AVB input is on. During backward locomotion, AVA input is on.}
\label{figBestEvolvedNetwork}
\end{figure}

The aim of this work is to explore the space of possible configurations of the repeating neural unit in the VNC that are capable of intrinsic network oscillations resembling those observed during forward and backward locomotion in the worm. However, the physiological parameters of this neural unit are largely unknown, including which chemical synapses are excitatory or inhibitory and their strengths, the strengths of the gap junctions, and the intrinsic physiological properties of the motorneurons. Therefore, we used an evolutionary algorithm to explore the configurations of the parameters that allow for an intrinsic oscillation in the neural unit so as to match the main features that have been experimentally observed in neural traces of forward and backward locomotion.  As the model does not include muscles or mechanical parts of the body, we focused on the patterns in the neural traces of the A- and B-class motorneurons as a proxy for evaluating forward and backward movement. Based on observations by Kawano et al.~\citep{Kawano2011}, we modeled command interneuron input to motoneurons as constant external inputs that could be switched ON or OFF (red and blue connections, Fig~\ref{figBestEvolvedNetwork}A). Finally, by running the evolutionary algorithm many times with different random seeds, we consider not just one possible configuration for the parameters of the neural unit, but as many variations as possible. As far as we are aware, this is the first attempt to explore the feasibility of CPGs in the VNC of \textit{C.~elegans}. 

%%%%%%%%%%%%  Figure best evolved network     %%%%%%%%%%%%

\section{Model and methods}

\subsubsection{Neuroanatomical unit}
We developed a model of a repeating neural unit in the ventral nerve cord (VNC) proposed in~\citep{Haspel2011} (Fig~\ref{fig_HaspelSegment}). The neural unit is based on an analysis of repeating patterns of connectivity along the VNC and a re-alignment of the motorneurons along the anteroposterior axis according to positions of the muscles they innervate. The repeating neural unit proposed in their analysis includes motorneurons: AS, DA, VA, DB, VB, DD, VD. We introduce a number of simplifying assumptions as a starting point for our investigation of the feasibility of VNC CPGs.  

First, our model assumes that cells of the same class are elecrophysiologically similar.  All neuron classes, except for DD and DB, comprise a pair of neurons: one anterior and one posterior. The neural unit is thus anterior-posterior symmetric in terms of cells by class. In this paper, we refer to anterior and posterior cells within a unit with the suffix `a' or `p' appended to the neuron's name (e.g., DAa and DAp for anterior and posterior DA cells within the neural unit).

Second, our model of the circuit introduces an anterior-posterior symmetry constraint on the connections (Fig~\ref{figBestEvolvedNetwork}A). The majority (79.4\%) of connections within the neural unit are anterior-posterior symmetric (black and brown connections, Fig~\ref{fig_HaspelSegment}). We assumed that the parameters of the connection (whether a connection was excitatory or inhibitory and its strength for a chemical synapse, and the strength for a gap junction) was anterior-posterior class symmetric. That is, for example, the connection between AS and DA was the same for the anterior part of the circuit as it was for the posterior half of the circuit. In order to further reduce the number of parameters in the model for our initial search, we also omitted the 8 connections that are not anterior-posterior symmetric (red connections, Fig~\ref{fig_HaspelSegment}). This simplification allows us to easily constrain the activity of cells from the same class within the unit to exhibit a similar pattern of activity as has been seen experimentally~\citep{Haspel2010}. 
%More broadly, the logic for the simplification in the context of our goal was as follows: If we could not find a CPG with the anterior-posterior symmetry constraint, then we would add back some of the asymmetry in the parameters and the connections gradually; if we found a CPG with the given constraints, than this would constitute a neuroanatomically-grounded instantiation of a CPG within the VNC.
%Based on preliminary experiments with the full asymmetrical circuit, we noticed that cells from the same class exhibited radically different patterns of activity. Therefore, 

Third, our model omits the DD motorneuron from the circuit. In the Haspel and O'Donovan unit (Fig~\ref{fig_HaspelSegment}), the DD motorneuron has only one outgoing connection (i.e., DD$\to$VD). This connection is not anterior-posterior symmetric (i.e., it connects to the anterior VD cell, but not to the posterior one). With the introduction of the anterior-posterior symmetry constraint on the connections, the DD neuron is left with no outgoing chemical synapse. Based again on preliminary experiments with the full asymmetrical circuit, we noticed that DD never showed a role in the generation of the oscillatory pattern. So as the starting point into our search for a potential VNC CPG, we omitted the DD motorneuron and its incoming connections (brown connections, Fig~\ref{fig_HaspelSegment}). While DD can play an important role in driving locomotion in the physical body, because of its neuromuscular junctions, due to its lack of outgoing connections within the neural unit, it is not likely to play a role in the generation of intrinsic network oscillations.

\subsubsection*{Neural model}
Following electrophysiological studies in \textit{C.~elegans}~\citep{Lockery2009,Mellem2008} and previous modeling efforts~\citep{Izquierdo2010,Izquierdo2013}, motorneurons were modeled as isopotential nodes with the ability to produce regenerative responses, according to:

\begin{eqnarray}
\label{eq:one}
\tau_{i} \frac{dy_{i}}{dt} = -y_{i} + \sum_{j = 1}^{N}{w_{ji}\sigma(y_{j} + \theta_{j})} + \sum_{k = 1}^{N}{g_{ki}(y_{k} - y_{i})} + I_{i}
\end{eqnarray}

\noindent where $y_i$ represent the membrane potential of the $i^{\text{th}}$ neuron relative to its resting potential, $\tau_i$ is the time constant, $w_{ji}$ corresponds to the synaptic weight from neuron $j$ to neuron $i$, and $g_{ki}$ as a conductance between cell $i$ and $j$ ($g_{ki}>0$). The model assumed chemical synapses release neurotransmitter tonically and that steady-state postsynaptic voltage is a sigmoidal function of presynaptic voltage~\citep{Kuramochi2017,Lindsay2011,Wicks1996}, $\sigma(x) = 1/(1 + e^{-x})$, where $\sigma(x)$ is the synaptic potential or output of the neuron. The chemical synapse has two parameters: $\theta_j$ is a bias term that shifts the range of sensitivity of the output function, and $w_{ji}$ represents the strength of the chemical synapse. 
%------- MINOR MODIFICATION based on Reviewer's Feedback, by E.I. -- OCT, 31, 2017 -----------%
When the bias term is a low value, the neuron is \textit{intrinsically inactive}: in the absence of excitatory stimuli the neuron output is off. When the bias term is a high value, the neuron is \textit{intrinsically active}: without external inhibition the neuron's synaptic release has a steady basal level.
%---------------------------------------------------------------------------------------------%
Unfortunately, there is no concrete evidence about the nature of electrical synapses in \textit{C. elegans}. Until more evidence is available, and in line with previous models~\citep{Wicks1996,Izquierdo2013,Kunert2017}, the model assumes electrical synapses are non-rectifying. The circuit was simulated using the Euler method with a time step of 0.0025.  

\begin{figure}[!ht]
\includegraphics[width=8.7cm]{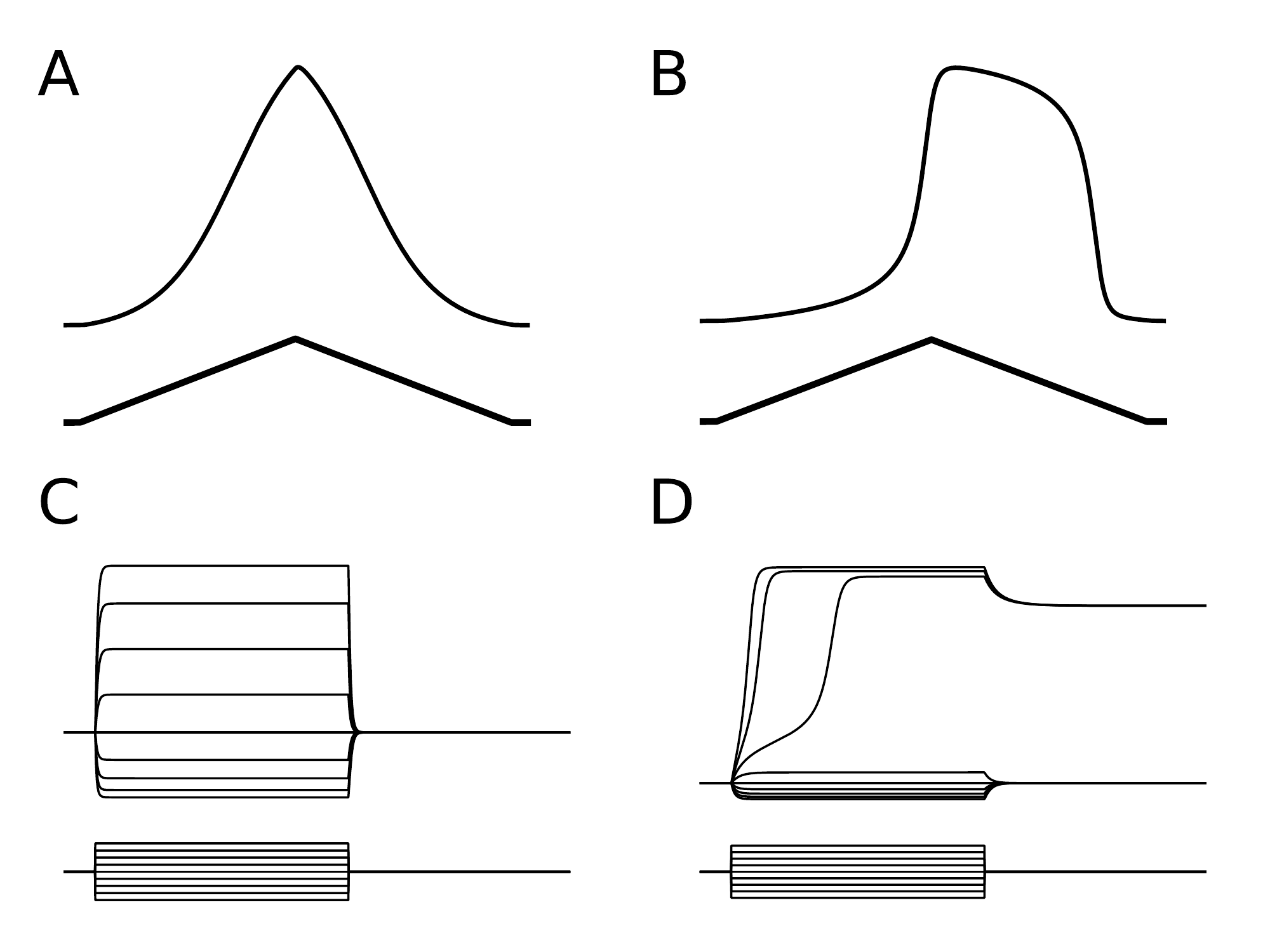}
\centering
\caption{{\bf Activity of model neuron resembles electrophysiological properties of \textit{C. elegans} neurons.}
We simulated the neural model in two different configurations to reproduce electrophysiological recordings in~\citep{Mellem2008}:  neuron AVA (left column) and neuron RMD (right column), as examples of passive and active dynamics, respectively. Parameters for the passive AVA-like neuron were $\theta$ = -1.8; $\tau$ = 0.03; $w$ = 0.6. Parameters for the active RMD-like neuron were $\theta$= -3.4; $\tau$ = 0.05; $w$ = 5.1. For each of the panels, the bottom trace represents the external input and the top trace represents the output of the neuron. Responses to a depolarizing current ramp from the passive neuron configuration [A] and the active response configuration [B] of the model. Responses to a current steps for the passive neuron configuration [C] and the active response configuration [D]. The active configuration reproduces the long-lived plateau potentials in response to depolarizing current steps observed in \textit{C. elegans} neurons~\citep{Mellem2008}.}
\label{fig_Electrophysiology}
\end{figure}

Our neural model has the capacity to  reproduce qualitatively the range of electrophysiological properties observed so far in \textit{C. elegans} neurons \citep{Lockery2009,Mellem2008} (Fig~\ref{fig_Electrophysiology}). The model can reproduce the passive activity that has been observed in some neurons (e.g., AVA): linear voltage response to depolarizing current ramps and a graded return to resting potential in response to current steps (Fig \ref{fig_Electrophysiology}A). Through the increase of the strength of the self-connection ($>4$, see \cite{Beer1995}), the model is also capable of reproducing the bistable potentials found in some neurons (e.g., RMD). The voltage response to depolarizing current ramps is initially linear, but then becomes regenerative, leading to a plateau potential (Fig \ref{fig_Electrophysiology}B). We also found that depolarizing current steps were sufficient to generate long-lived plateau potentials, as in RMD neurons. On cessation of the current step, the voltage relaxed to a different steady-state value from the initial resting potential (Fig \ref{fig_Electrophysiology}D), also as in RMD neurons.

%%%%%%%%%%%%  Figure Electrophysiology     %%%%%%%%%%%%

\subsubsection*{Interneuron inputs}
Interneuron control of locomotion direction depends on a relatively complex interaction between electrical and chemical synapses~\citep{Kawano2011}. While deletion of no single command neuron disrupts forward or backward locomotion entirely, deletion of AVA and AVB produce the most significant differences in behavior~\citep{Chalfie1985}. During forward locomotion, AVB activity is active, while AVA is suppressed. During backward locomotion, the opposite is true: AVB is suppressed, while AVA is active. Although modeling the command interneuron circuit with all its synapses is outside the scope of this paper, the main role of the AVB class is to influence the B-class motorneuron to turn on and off forward locomotion, while the main role of AVA is to influence the A-class motorneuron to turn on and off backward locomotion ~\citep{Zheng1999, Kawano2011, qi2012photo, Rakowski2013}. As we are not modeling the body, and therefore speed, we do not take into account results about the influence of the slope of AVA or the basal level effect on worm speed~\citep{kato2015global}. Accordingly, we modeled interneuron control of locomotion as a binary external input over A- and B- class motorneurons. During the simulation of the circuit, we refer to forward locomotion when the B-class motorneuron receives external input from AVB and the A-class motorneuron does not receive input from AVA. We refer to backward locomotion when the B-class motorneuron receives no external input from AVB, and the A-class receives external input from AVA.

\subsubsection*{Evolutionary optimization}
The parameters of the model were evolved using a genetic algorithm. The optimization algorithm was run for populations of 1000 individuals. Each individual encoded the 34 unknown parameters of the model. These unknown parameters included those that determined the physiology of each of the six neuron classes in the model: self-weight, bias and time-constant, all together 18 parameters. It also included the weights and polarities of the chemical synapses and gap junctions. The neural unit has a total of 20 chemical synapses and 6 gap junctions. We imposed an anterior-posterior symmetry in the sign and strength of the chemical and electrical connections, reducing the total number of unknown parameters to 10 for chemical and 4 for electrical synapses. Finally, the input from the command interneurons were modeled with 2 parameters: the strength from AVB to B-class motorneurons and AVA to A-class motoneurons. The range for neurons time-constant was set to [0.05, 2]. The range for biases, self-weights, chemical synapses, and interneuron input was set to [-20, 20]. The range for electrical synapses was set to [0, 2.5]. We ran 1000 evolutionary algorithms with different random seeds. Each evolutionary experiment was run for 1000 generations.

\subsubsection*{Fitness evaluation}
Fitness was evaluated in a simulated forward and backward assay. At the start of each assay, a circuit was presented with the forward condition (AVB input), the transients were allowed to pass for 6 units of time, and then the neural traces were evaluated for the following 20 units of time. Then the circuit's state was reset and presented with the backward condition (AVA input). The procedure was repeated similarly, transients were allowed to pass and then neural traces were evaluated. For each locomotion direction, neural traces were evaluated using three components that capture the main three features of the \textit{in-vivo} calcium recordings for these motorneurons during forward and backward locomotion~\citep{Kawano2011,Faumont2011}. Total fitness for an individual corresponded to the multiplication of the three components: 
%$\begin{eqnarray}
%\label{eq:two}
$F = F_{1}\, *\, F_{2}\, *\, F_{3}$\
%$\end{eqnarray}

\subsubsection*{Oscillation criterion} 

Intrinsic network oscillations were measured by evaluating the total change in the derivative of the output in the dominant class of neurons:
\begin{eqnarray}
\label{eq:three}
F_1 = \prod_{Y}{S_Y} 
\end{eqnarray}
where
\begin{eqnarray}
\label{eq:five}
% 	O_N = \frac{2}{A*T}\int_0^T{\big|\frac{dN}{dt}\big|}dt
	S_Y = \frac{2}{A*T}\int_0^T{\left\lvert\frac{dO_Y}{dt}\right\rvert dt}
\end{eqnarray}
\noindent where $A$ corresponds to an optimal oscillation amplitude covering roughly one third of the full output range ($A=0.3$). Oscillations in this limited range match what has been observed experimentally~\citep{Kawano2011}, which ultimately allows for one class of motorneurons to show dominance over the other one, by oscillating in the top third or the bottom third of the full output range. $T$ represents the simulation time length ($T=20$ units of time). $dO_Y$ represents the rate of change of the output of neurons. $S_Y$ is capped at 1. The set of dominant neurons, Y, is \{DB, VBa, VBp\} for forward locomotion and \{DAa, DAp, VAa, VAp\} for backward locomotion.

\subsubsection*{Phase criterion} 

Dorsoventral out-of-phase oscillations were evaluated by measuring the difference in sign between the derivatives of the dorsal and ventral outputs, according to:

\begin{eqnarray}
\label{eq:six}
F_{2} =  \prod_{(V, D)}{\left(1 - \frac{1}{2T}\int_0^T{\big|sgn(dO_V) + sgn(dO_D)\big|\,dt}\right)}
\end{eqnarray}
\noindent where $dO_V$ and $dO_D$ represent the rate of change of the output of ventral and dorsal motorneurons, from the set \{(VBa, DB), (VBp, DB)\} for forward locomotion and \{(VAa, DAa), (VAp, DAp)\} for backward locomotion.

\subsubsection*{Dominance criterion} 

The dominance of A- and B- classes during forward and backward locomotion was assessed using three components: the minimum output value in the dominant neuron class, the maximum output value in non-dominant neuron class, and the amplitude of the oscillation in the dominant neuron class: 

\begin{eqnarray}
\label{eq:seven}
F_3 = \prod_{Y} {f(m_Y, (1-A))}\, *\, \prod_{X} {f(M_X, A)}\, *\, \prod_{Y} {f(A_Y, A)} \
\end{eqnarray}
\noindent where $m_Y$ represents the minimum value in the trace of the output from neurons in the dominant class ($m_Y=\min\limits_{t}O_Y(t)$), $M_X$ represents the maximum value in the trace of the output from neurons in the non-dominant class ($M_X=\max\limits_{t}O_X(t)$), and $A_Y$ represents the amplitude of the oscillation in the trace of the output from neurons in the dominant class, as given by the difference between the minimum value and the maximum value ($A_Y=\max\limits_{t}O_Y(t) - \min\limits_{t}O_Y(t)$). 
During forward locomotion, $Y$ = \{DB, VBa, VBp\} and $X$ = \{DAa, DAp, VAa, VAp\}. The opposite is true for backward locomotion. $A$ corresponds to the optimal oscillation amplitude (see previous subsection). We use a normalization function $f$, which is a non zero, smooth function in the range [0,1], maximal when $x = x_0$:

\begin{eqnarray}
\label{eq:eleven}
f(x, x_0) = 0.1  +  0.9 \,\,\, \left(\frac{x}{x_0}\right) \,\, e^{(1 - x/x_0)}
\end{eqnarray}

\subsubsection*{Post-search filtering}
When we analyzed the ensemble of best individuals, we noticed that some of the solutions showed damped oscillations.  To remove such solutions, we simulated the ensemble for a much longer time period (3000 units of time) and reevaluated them under the same fitness function. For the analysis we considered only circuits with stable oscillations. 

\subsubsection*{Ablations}
\label{sec:ablation}

In order to understand the role of some neuron in a circuit's operation, it is common to eliminate the neuron from the circuit entirely through ablations. Unfortunately, ablating a neuron conflates two somewhat different effects. First, the operation of a postsynaptic neuron may depend on the temporal details of the signal it receives from a presynaptic neuron. Second, a presynaptic neuron's tonic baseline of activity might be necessary to maintain a postsynaptic neuron's activity in its appropriate operating range. In order to distinguish between these two different effects, we examine the extent to which replacing the signal in one connection with a tonic input is sufficient to maintain the normal operation of the circuit. 

In our analysis of the evolved CPGs, we systematically study the participation of each connection by evaluating the circuit's performance when the connection is substituted by a constant input. As the constant input that better substitutes the dynamic input for a given connection is unknown, we evaluated each connection using a batch of 1000 different constant inputs when studying chemical synapses and 2000 values for electrical synapses. From this batch of simulations, we selected the best fitness achieved in forward and backward locomotion independently. Finding a tonic input that can substitute a connection with little or no effect to the circuit's performance suggests that the presynaptic neuron is keeping the postsynaptic neuron within its operating range, but that there is no temporal detail in the signal being transmitted. Not finding a tonic input that allows the circuit to perform normally suggests that the presynaptic neuron's dynamical activity is crucial for the operation of the circuit. When a neuron has no effect on any of its postsynaptic neurons, then that neuron can be said to not play a role in the circuit.

% Results and Discussion can be combined.
\section*{Results}
\subsection*{VNC neural unit can intrinsically generate locomotion-like oscillations}

The \textit{C.~elegans} locomotion pattern can be characterized by three features that any putative CPG must exhibit: (1) \textit{oscillation criterion}: the A- and B-class motorneurons oscillate~\citep{Haspel2010,Kawano2011,Faumont2011}; (2) \textit{phase criterion}: the dorsal and ventral oscillations in each class of cells occur in antiphase in order to drive the alternating dorsoventral bends responsible for locomotion~\citep{Kawano2011,Faumont2011}; (3) \textit{dominance criterion}: B-class activity exceeds A-class activity during forward locomotion, whereas A-class activity exceeds B-class activity during backward locomotion~\citep{Haspel2010,Kawano2011}. Each of these features become a term in the measure of locomotion performance that we optimized. In total, 1000 optimizations with different initial random seeds were run. The best-performing circuit obtained from each run forms a solution ensemble whose properties we study in this paper.

\begin{figure}[!ht]
\includegraphics[width=8.7cm]{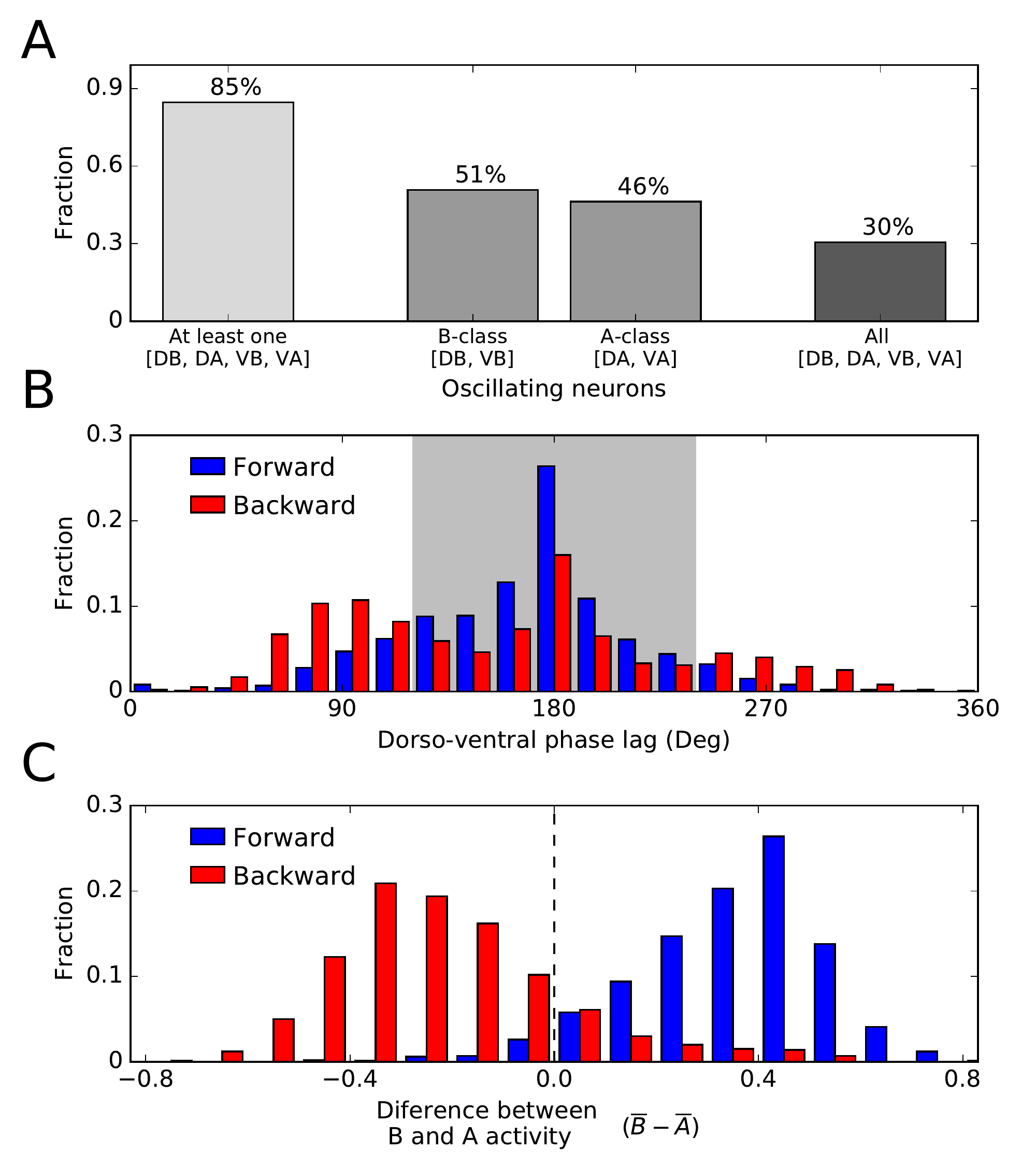}
\centering
\caption{{\bf Performance of the ensemble on the individual locomotion criteria.}
[A]~Oscillation criterion. Fraction of solutions that accomplish oscillations in one or more neurons in the network. B-class neurons were evaluated during forward locomotion and A-class during backward. 30\% of all solutions showed oscillations in all relevant motorneurons. [B]~Phase criterion. Dorsoventral phase lag between B-class neurons for forward locomotion and A-class neurons during backward locomotion. The majority of solutions oscillate in dorsoventral antiphase. [C]~Transition criterion. Distribution of the differences in mean activity between B- and A-class neurons during forward and backward locomotion. In the majority of solutions, the A-class dominates during backward and the B-class dominates during forward.}
\label{fig_3Criteria}
\end{figure}

The breakdown of the ensemble with respect to the three individual criteria is as follows. With respect to the oscillation criterion, over 80\% of the ensemble exhibited oscillations in at least one neuron (Fig~\ref{fig_3Criteria}A). Specifically, 511 solutions fulfilled the criteria for forward locomotion, 463 solutions fulfilled the criteria for backward locomotion, and a total of 308 solutions achieved oscillations for both forward and backward locomotion. With respect to the phase criterion, over 78\% of the circuits in the ensemble exhibited DB/VB antiphase during forward locomotion, whereas over 46\% of the circuit exhibited antiphase in DA/VA neurons  during backward locomotion (Fig~\ref{fig_3Criteria}B). Finally, with respect to the dominance criterion, 81\% of the solutions exhibited the proper relative magnitude between the A- and B-class neurons, with B-class activity dominating during forward locomotion and A-class activity dominating during backward locomotion (Fig~\ref{fig_3Criteria}C).

%%%%%%%%%%%%  Figure 3 Criteria     %%%%%%%%%%%%

\begin{figure}[!ht]
\includegraphics[width=9.7cm]{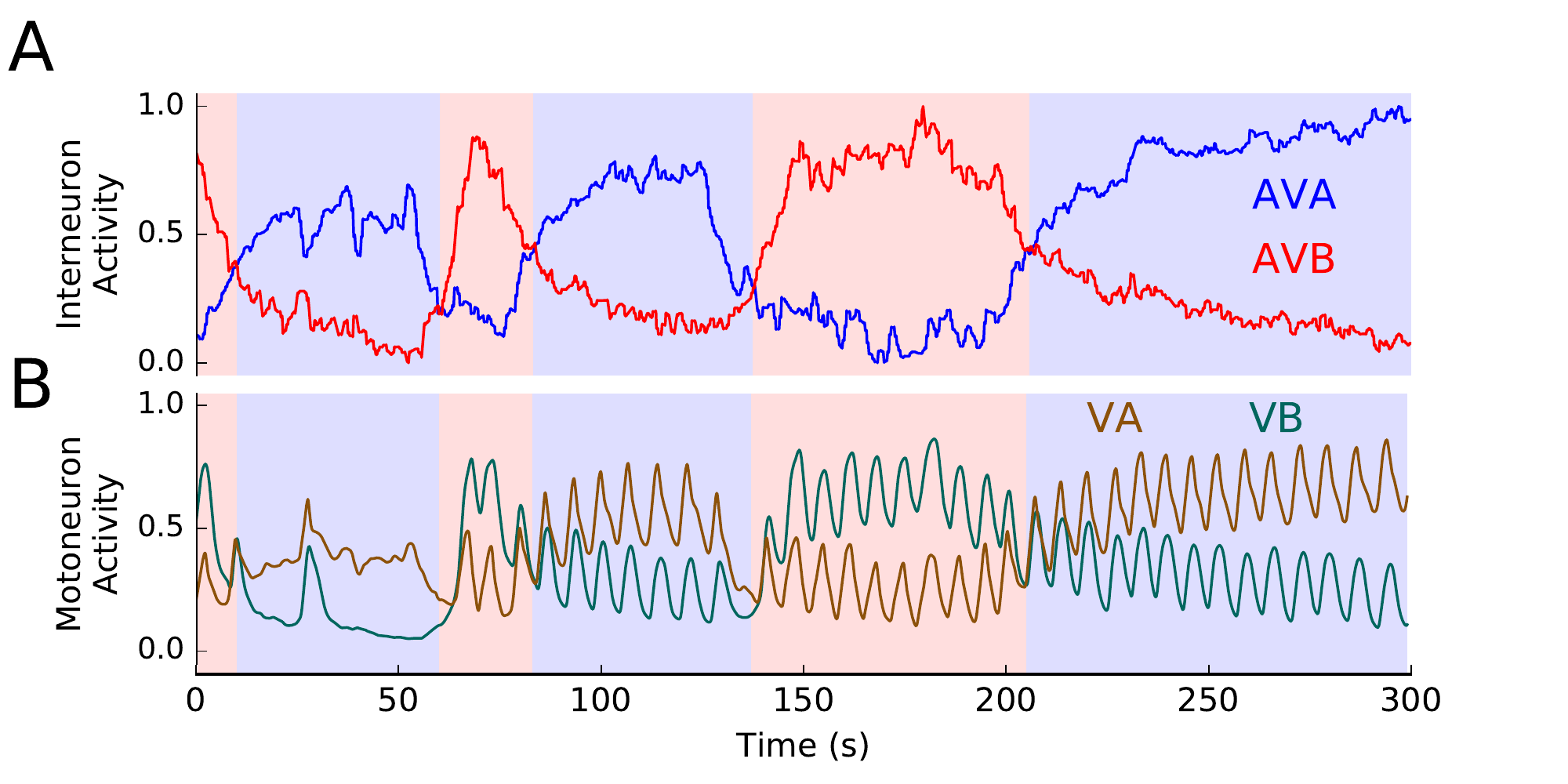}
\centering
\caption{{\bf Activity of best evolved circuit during realistic stimulation by AVA and AVB command interneurons.}
[A]~Activity of command interneurons recorded in~\citep{Kawano2011}. Traces were normalized. Background color indicates forward and backward dominance depending on relative activations of AVA and AVB. [B]~Traces from VA and VB neurons in the simulated circuit when stimulated by the traces in [A].}
\label{figInterneuronInput}
\end{figure}

We found that over 11\% of the circuits in the ensemble satisfied all three criteria, demonstrating that the repeating neural unit in the VNC is indeed capable of intrinsically generating worm-like locomotory oscillations that can be switched between a forward and backward mode by appropriate AVA and AVB command neuron input (Fig~\ref{figBestEvolvedNetwork}). Interestingly, when the model circuit is driven by realistic AVA/AVB input taken from calcium imagining of freely moving worms (see Fig~1F in~\citep{Kawano2011}, reproduced in Fig~\ref{figInterneuronInput}A), it produces realistic-looking motor output (compare Fig~\ref{figInterneuronInput}B to Fig~2A in~\citep{Kawano2011}). Specifically, the model maintains all three locomotion criteria throughout the trial, with smooth transitions in dominance between A- and B- class motorneurons that correlate with AVA and AVB activity.

%%%%%%%%%%%%  Figure best network fed with interneuron recordings     %%%%%%%%%%%%

\subsection*{Operation of the Best Evolved Circuit}
Given that the possibility of a CPG in the VNC has not been seriously considered in the literature, it is interesting to examine how this feat is achieved. Accordingly, we next analyze the circuit with the best overall performance in the ensemble (Fig~\ref{figBestEvolvedNetwork}A). This circuit ranked 1\textsuperscript{st} on the oscillation criterion, 12\textsuperscript{th} on the phase criterion, and 11\textsuperscript{th} on the dominance criterion. When we simulated neural traces of this circuit for forward and backward locomotion, we observed traces that matched all the main features~\citep{Haspel2010,Kawano2011,Faumont2011} (Fig~\ref{figBestEvolvedNetwork}B): (1) A- and B-class motorneurons oscillate; (2) Dorsal and ventral motorneurons oscillate in antiphase; and (3) B-class baseline oscillatory activity is higher than A-class baseline oscillatory activity during forward locomotion, and the opposite is true during backward locomotion. Note that this third feature is achieved despite the activity of DB only slightly decreasing during backward locomotion. In this section, we characterize the operation of this best evolved circuit.  First, we characterize a dorsal subcircuit that serves as the core oscillator driving the locomotion pattern. We then determine how these dorsal oscillations propagate to the ventral motorneurons to produce antiphase oscillations. Finally, we show how the remaining connections in the circuit fine-tune the features of the observed pattern.

\subsubsection*{A dorsal core subcircuit capable of oscillations}

\begin{figure}[!ht]
\includegraphics[width=13.2cm]{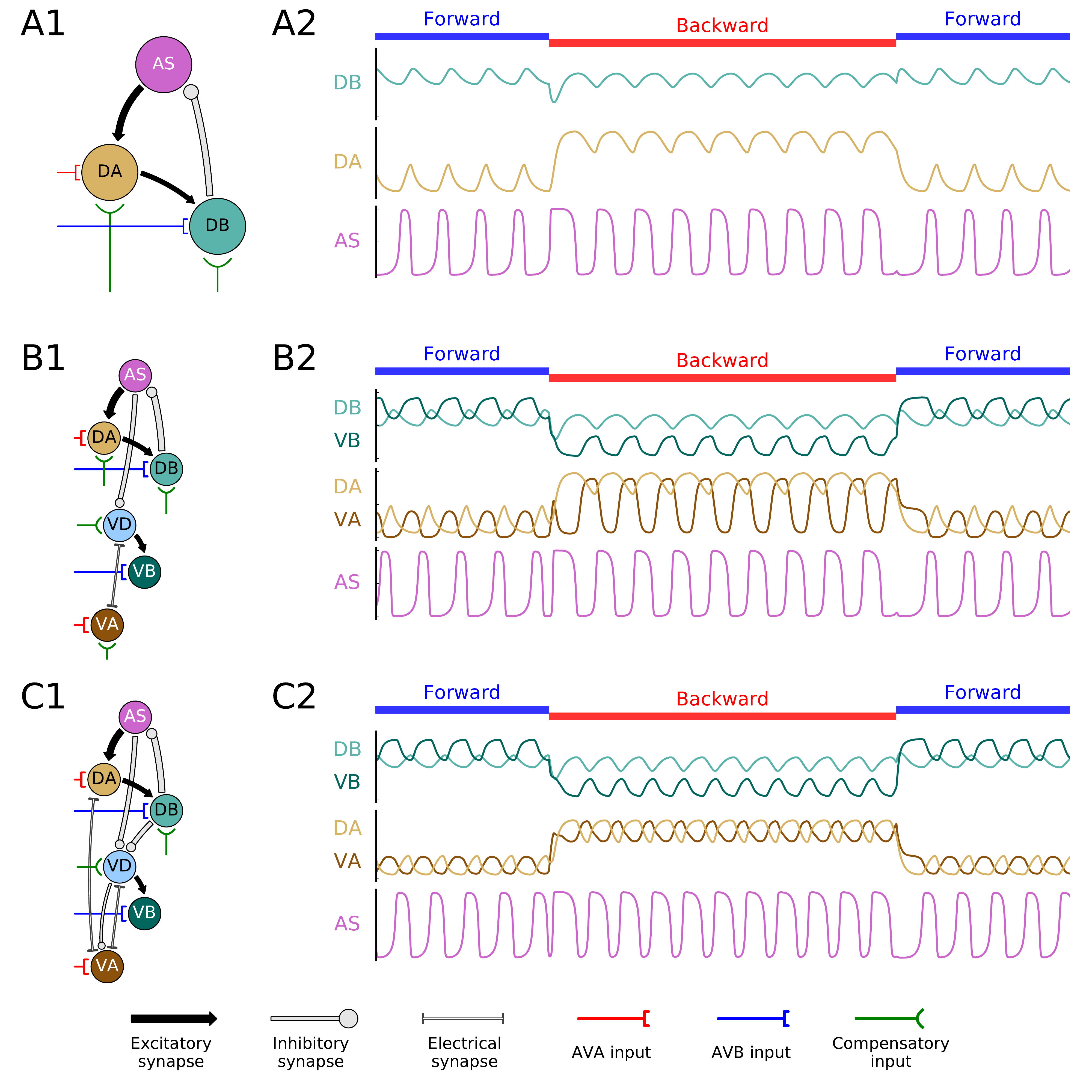}
\centering
\caption{{\bf Functional analysis of the best evolved circuit.}
[A1]~Dorsal core subcircuit. Minimal CPG comprises all three neurons that innervate dorsal muscles. Legend for the synaptic connections at the bottom of the figure. Connection width is proportional to synaptic strength. [A2]~Output traces of the dorsal core subcircuit under forward and backward external inputs. Traces maintain shift in relative activation of B/A class neurons. Input from neurons AVB and AVA are shown with blue and red bars respectively. [B1]~Minimal functional subcircuit. All features of the locomotion pattern are present when VD, VB, and VA are added to the dorsal core subcircuit. [B2]~Simulated neural traces for the minimal functional subcircuit. Main output difference compared to the full network is the increase in amplitude in VA neuron during backward locomotion. [C1]~Extended minimal circuit. [C2]~Simulated neural traces for the extended minimal circuit.}
\label{figFromDorsalToVentral}
\end{figure}

How does the best circuit in the ensemble produce oscillations? Since we did not include pacemaker neurons in our model, the circuit must contain a network oscillator, that is, a sub-circuit of neurons that collectively generate a rhythmic pattern through their interactions. Systematically ablating classes of neurons revealed a dorsal core subcircuit consisting of AS-, DA-, and DB- classes capable of generating oscillations (top half of the circuit in Fig~\ref{figBestEvolvedNetwork}A). No other oscillatory subset of neuron classes was identified. In fact, further analysis demonstrated that either half of this dorsal subcircuit could oscillate in isolation, as long as appropriate tonic inputs to DA and DB were substituted for the missing neurons in order to maintain these neurons within their operating ranges (Fig~\ref{figFromDorsalToVentral}A1). Interestingly, this dorsal core subcircuit includes the two classes of motorneurons that have been most strongly implicated in locomotion (A- and B-) \citep{Chalfie1985, Haspel2010}, as well as another class (AS) that has not yet been well studied experimentally.

%%%%%%%%%%%%  Figure dorsal cord and spreading to ventral     %%%%%%%%%%%%

\begin{figure}[!ht]
\includegraphics[width=13.2cm]{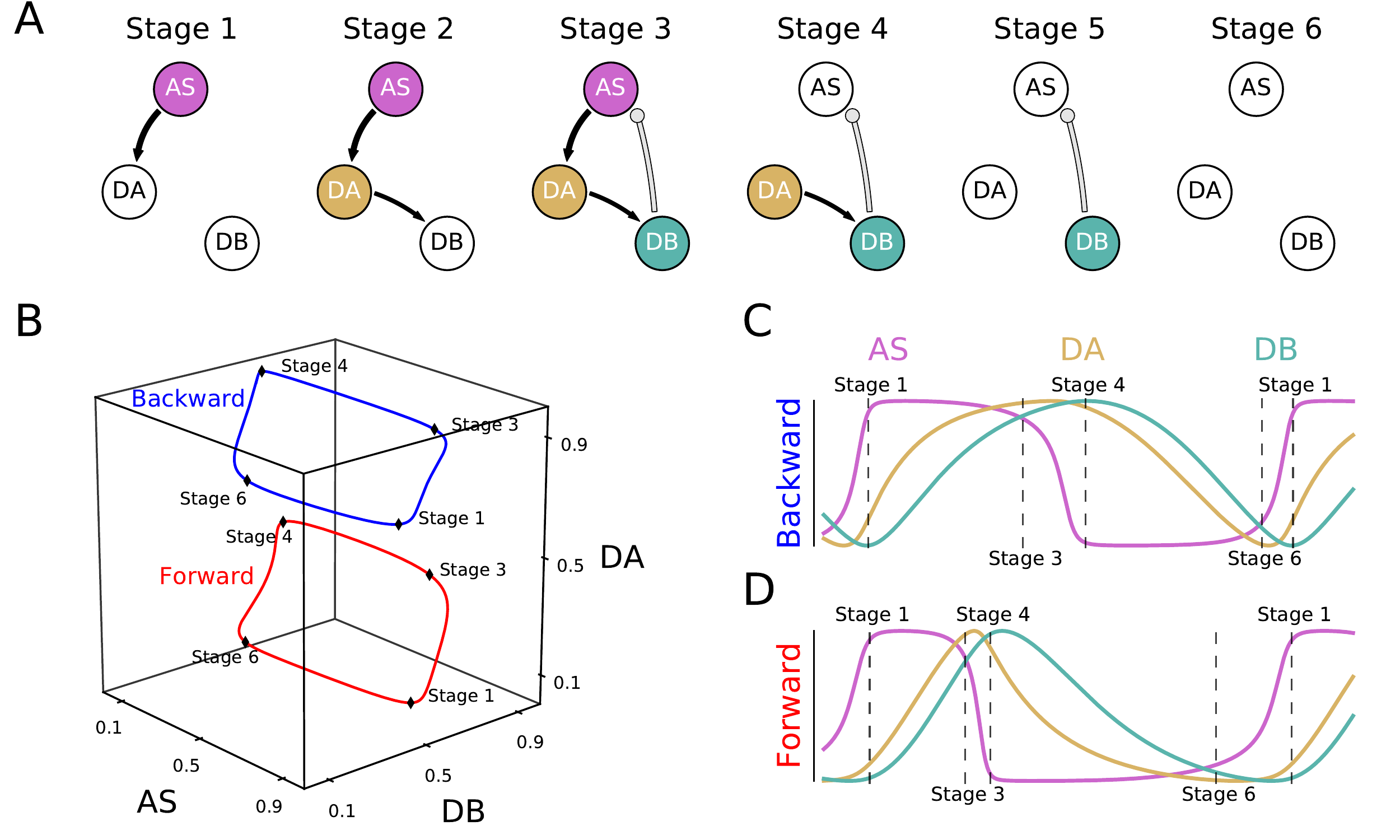}
\centering
\caption{{\bf Dorsal core operation.}
[A] The cyclic operation of the dorsal core subcircuit can be broken down into 6 stages. Each stage represents a ``quasi-stable'' configuration of the subcircuit. Active neurons are colored; inactive neurons are shown in white. Active connections are shown; inactive connections are not shown. Dorsal core subcircuit operates similarly during forward and backward locomotion. [B] Phase plot of dorsal core neurons during backward and forward simulations. Stages 1, 3, 4 and 6 are shown with disks, according to panel [A]. Dorsal core neural traces for backward [C] and forward [D]. Neural traces normalized. Vertical lines mark stages 1, 3, 4 and 6 in the oscillation cycle. Although the timing of stages is the different for forward and backward, the sequence is the same.}
\label{figDorsalCore}
\end{figure}

The operation of this dorsal core subcircuit is straightforward to explain. AS is bistable (see Fig \ref{fig_Electrophysiology}B) and intrinsically active
% ------- MODIFICATION DUE TO LAST REVIEW by E.I. OCT 31, 2017
% {\color{red}(i.e., in the absence of input it adopts its maximum value)}, 
% whereas DA and DB are monostable and intrinsically inactive
% {\color{red}(they adopt their minimum values in the absence of input)}. 
% ----------------------------------------------------------------------
At the beginning of a locomotion cycle, AS is active and both DA and DB are inactive (Fig~\ref{figDorsalCore}A, stage 1). Since AS excites DA, activity in the former begins to activate the latter (stage 2), which in turn excites DB, activating it as well (stage 3). But, since DB inhibits AS, its activity switches AS off (stage 4), removing the source of excitation that maintains DA and DB. As activity in DA fades (stage 5), so does activity in DB (stage 6). As AS is released from inhibition the cycle repeats, creating a CPG.

%%%%%%%%%%%%  Figure Stage dorsal core cycle     %%%%%%%%%%%%

The operation of the dorsal core subcircuit is qualitatively the same for forward and backward locomotion (Fig~\ref{figDorsalCore}B, C, D). Although there are some small changes in the dynamics of the A- and B- class motorneurons as a result of the change of input from the command motorneurons, for the purposes of our analysis, the primary feature that changes is the level of their baseline oscillation. Both AVA and AVB evolved to excite A- and B-class motorneurons. During forward locomotion, input from AVB into B-class motorneurons produces a slight increase in the mean level of activity of B-class neurons (right shift in the location of the red limit cycle, Fig~\ref{figDorsalCore}B), dominating the activity of the A-class neurons. The opposite is true during backward locomotion. Input from AVA into the A-class motorneurons increases the mean level of activity in the A-class activity (upward shift in the location of the blue limit cycle, Fig~\ref{figDorsalCore}B), which ends up dominating the activity of the B-class neurons.

\subsubsection*{Dorsoventral coordination}

How does the dorsal oscillation spread to the ventral motorneurons in an antiphase manner? Systematic ablation of the connections from dorsal to ventral neurons demonstrates that the primary route of transmission is via the inhibitory connection from AS to VD. Like AS, VD is also bistable and intrinsically active. As AS is switched on and off by its interactions with DA and DB, it switches VD off and on, respectively, in an antiphase manner. This antiphase activity in VD is in turn propagated to VA (VD$\vdashv$VA) and to VB (VD$\to$VB). Thus, the minimal functional subcircuit is comprised of the dorsal motorneurons AS, DA and DB and the ventral motorneurons VD, VA and VB along with the connections shown in Fig~\ref{figFromDorsalToVentral}B. Note that, once again, tonic compensatory input to VD and VA are required in order for these neurons to remain within the appropriate operating range.

\subsubsection*{Fine-tuning the pattern}

The only significant difference remaining between the oscillation pattern of the minimal functional subcircuit (Fig~\ref{figFromDorsalToVentral}B) and that of the full circuit (Fig~\ref{figBestEvolvedNetwork}) involves VA. As in the worm~\citep{Haspel2010,Kawano2011}, during backward locomotion in the full circuit the minimum activity of VA remains quite high. However, in the minimal functional subcircuit, the minimum VA activity falls to almost zero under the same conditions (Fig~\ref{figFromDorsalToVentral}B2).  We found that we could restore the amplitude of VA activity only by including the gap junction DA$\vdashv$VA and the inhibitory chemical connection VD$\to$VA. Interestingly, the shape and amplitude of the VA oscillation cannot be simultaneously maintained by tonic substitution alone; phasic interaction is required. With these connections (Fig~\ref{figFromDorsalToVentral}C), the minimal functional subcircuit is capable of producing a pattern that satisfies all three locomotion criteria and is nearly identical to that of the full circuit. Finally, the contributions of the rest of the components in the full circuit were minor (see white disks in Fig~\ref{fig_EnsembleAblation}). Substituting chemical synapses VA$\to$VD and DA$\to$VD for a compensatory tonic input reduced locomotion fitness to 93\% and 99\%, respectively. Substituting gap junctions VD$\vdashv$VD and VB$\vdashv$VB did not reduce locomotion fitness at all.
 
\subsection*{Variations in the ensemble}

The analysis in the previous section demonstrates in detail one way in which a CPG could operate in the VNC of \textit{C.~elegans}. However, the experimental constraints are sufficiently weak at present that many other circuit configurations may also be consistent with the observed neuroanatomy, making it difficult to draw general conclusions from this single example. Fortunately, our ensemble of 110 solutions (i.e., 11\% out of the 1000 solutions obtained from optimization runs with different initial random seeds that satisfy all three locomotion criteria) represents a significant sample of the space of possibilities. In this section, we examine the operation of this entire set of solutions, with a particular focus on similarities and difference with the best evolved circuit.

\subsubsection*{Dorsal core subcircuit}  

In the best circuit, the dorsal motorneurons AS, DA and DB serve as the core pattern-generating subcircuit. How common is this core subcircuit in the ensemble? We found that this same set of neurons drives the locomotion pattern in 108 out of the 110 solutions (Fig~\ref{fig_EnsembleAblation}A). As in the best circuit, AS was bistable in 83.6\% of the solutions, whereas the A-class and B-class motorneurons were only bistable 3.6\% of the time. 

For the following analysis, we focused exclusively on the 108 solutions with a dorsal core subcircuit. Of the two solutions that did not utilize the dorsal core subcircuit to generate oscillations, one was driven by a ventral oscillator comprised of the pair VA-VD (ranked 32 out of 108) and the other circuit required all neurons except for AS to produce oscillations (ranked 91 out of 108). This suggests that, although the dorsal core subcircuit was an easily available solution, it is not the only possible way of producing oscillatory activity in the VNC. However, as these two solutions were atypical, and since the main purpose of our contribution is to demonstrate the possibility of a CPG in the VNC, we will not analyze them in any more detail.

%%%%%%%%%%%%  Figure Dorsal core classes     %%%%%%%%%%%%
\begin{figure}[!ht]
\includegraphics[width=9.7cm]{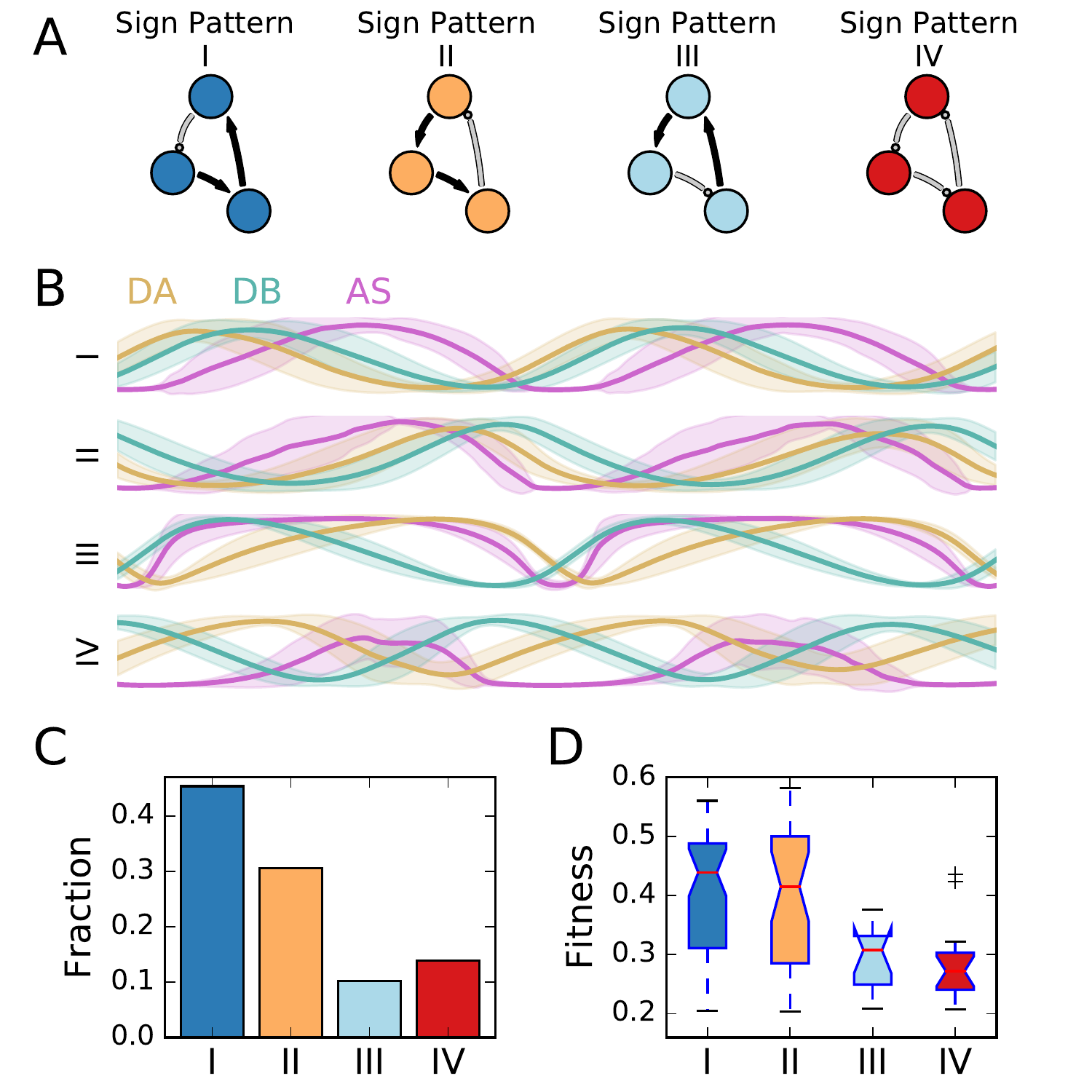}
\centering
\caption{{\bf Sign patterns found in the ensemble for the dorsal core subcircuit.}
[A]~Different patterns of signs of the connections in the dorsal core. Black arrows represent excitatory chemical synapses. Circle-ending connections represent inhibitory chemical synapses. [B]~Solutions in each group show a characteristic oscillatory pattern. Traces correspond to average $\pm$standard deviation of the normalized neuronal outputs. Two oscillation cycles are shown. [C]~Proportion of the total solutions in the ensemble with each of the different dorsal core sign patterns. Sign patterns I and II represent over 75\% of the solutions. [D]~Fitness for the solutions according to their dorsal core sign pattern. Solutions with sign patterns I and II comprise the best performing solutions.}
\label{figEnsembleClasses}
\end{figure}
  
However, despite the near-universal appearance of dorsal core oscillators, we do find variations in the pattern of excitatory and inhibitory synapses within the ensemble. Four different sign patterns of dorsal core subcircuits were identified (Fig~\ref{figEnsembleClasses}A), with the best circuit having sign pattern II. Sign patterns I, II and III all contain two excitatory synapses and one inhibitory synapse, differing only in the placement of the latter. Sign pattern IV, in contrast, contains only inhibitory synapses. The four group of solutions exhibit neural activity patterns which are consistent within each group but differ across groups (Fig~\ref{figEnsembleClasses}B). Note that DA and DB (yellow and green traces) oscillate nearly in phase in solutions with sign patterns I and II (in which the DA$\to$DB connection is excitatory), whereas they oscillate out of phase in solutions with sign patterns III and IV (in which the same connection is inhibitory).

%%%%%%%%%%%%  Figure Ablations     %%%%%%%%%%%%
\begin{figure}[!ht]
\includegraphics[width=11.4cm]{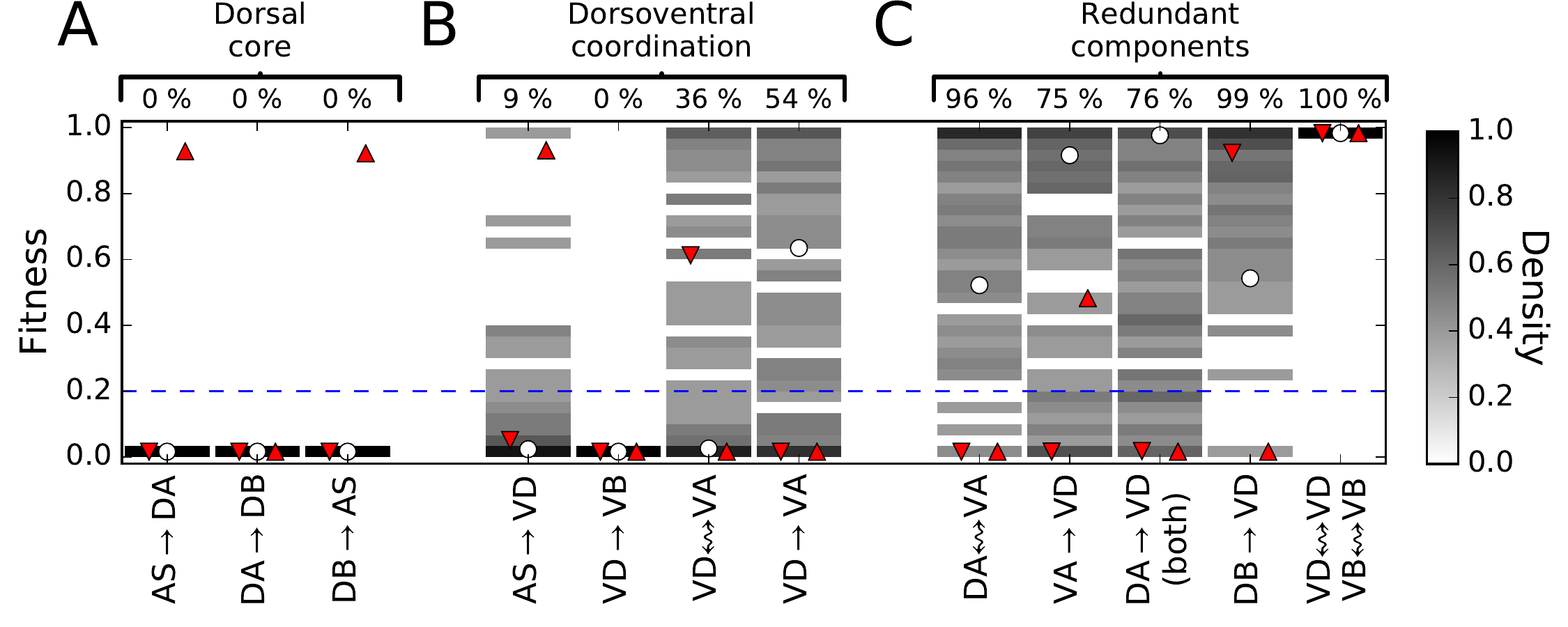}
\centering
\caption{{\bf Ablation experiments.}
Vertical histograms show the normalized fitness distribution among the 108 solutions that evolved a dorsal core oscillator while each of the connections is exchanged for a compensatory input. The dashed line marks a minimum level of fitness of 20\%. This may seem low, but the fitness is multiplicative over 8 different component, such that on average each different components of fitness has above 80\% performance. The percentage of solutions that maintain a fitness above this level is indicated at the top of the respective column.  The white disk corresponds to the best evolved circuit. The red triangles correspond to the two solutions that did not evolve a dorsal core oscillator. The connections are divided into: [A]~connections present in the dorsal core subcircuit; [B]~connections in the minimal functional subcircuit; [C]~connections responsible for minor contributions.}
\label{fig_EnsembleAblation}
\end{figure}
 
These four sign patterns were not equally represented across the ensemble (Fig~\ref{figEnsembleClasses}C). Sign patterns I and II occur much more frequently than the other two, accounting for 75.9\% of the solutions altogether. In addition to being the most represented, solutions with sign patterns I and II had on average higher fitness than solutions with sign patterns III and IV  (Fig~\ref{figEnsembleClasses}D). The other four logically possible combinations of excitation and inhibition among dorsal motorneurons were not observed in the ensemble.
 
\subsubsection*{Dorsoventral coordination}

\begin{figure}[!ht]
\includegraphics[width=8.7cm]{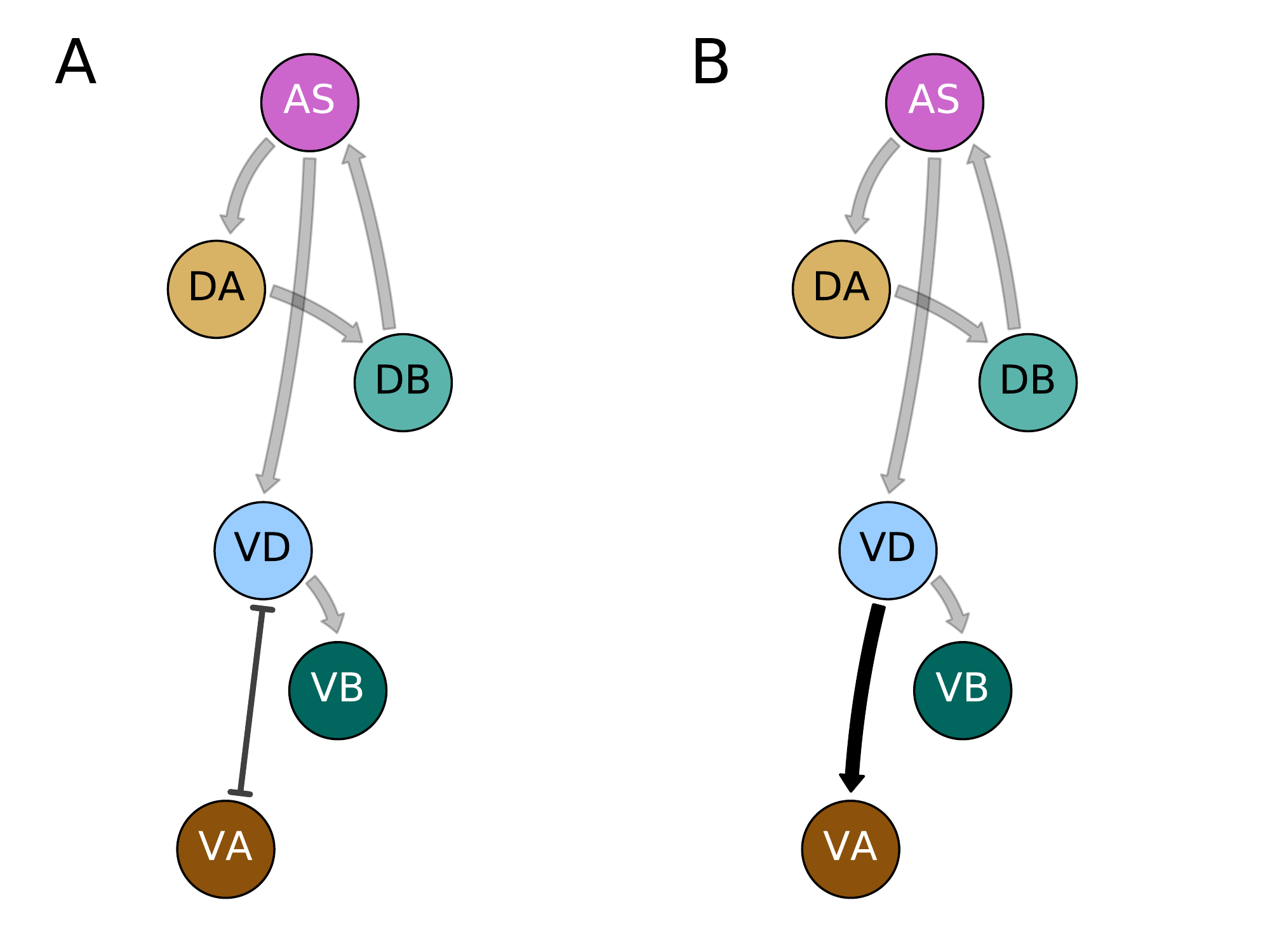}
\centering
\caption{{\bf Two forms of the minimal functional circuits found in the ensemble.}
Common connections shown in gray. Unique connection shown in black. In both forms, the dorsal motorneurons AS, DA and DB serve as the core pattern-generating subcircuit, dorsal oscillations propagate to ventral motorneurons in an antiphasic manner via the chemical synapse AS$\to$VD, and from there it propagates to VB via the chemical synapse VD$\to$VB. In one form of the solution, the information gets passed on to VA via a gap junction~[A] and in others via a chemical synapse~[B]. 
}
\label{fig_TwoMinimalFunctionalCircuits}
\end{figure}

In the best circuit, dorsal oscillations propagate to ventral motorneurons in an antiphasic manner via the chemical synapse AS$\to$VD, and then from there to VA and VB via VD$\vdashv$VA and VD$\to$VB, respectively. How common were these patterns in the ensemble? We found that the AS$\to$VD chemical synapse was necessary for dorsoventral coordination in 90\% of the solutions and sufficient in 78\% (Fig~\ref{fig_EnsembleAblation}B). Unsurprisingly, the VD$\to$VB chemical synapse was always essential, since no other path to VB exists in the repeating neural unit. However, two paths exist to VA: through the VD$\vdashv$VA gap junction and through the VD$\to$VA chemical synapse. As in the best circuit, 64\% of the solutions relied on the electrical connection, whereas the chemical synapse was necessary 46\% of the time (some relied on both). Finally, VD was bistable in only 20\% of the solutions, suggesting this feature is not essential. Therefore, the minimal functional subcircuit takes two forms in the ensemble, with the only difference being the type of connection from VD to VA (Fig~\ref{fig_TwoMinimalFunctionalCircuits}).

%%%%%%%%%%%%  Two most common functional circuits     %%%%%%%%%%%%

\subsubsection*{Fine-tuning the pattern}

In the best circuit, the connections VD$\to$VA and DA$\vdashv$VA played an important role in fine-tuning the locomotion pattern. In the ensemble, 46\% of the solutions also depended on the VD$\to$VA chemical synapse for fine-tuning. However, 96\% of these solutions performed well without the DA$\vdashv$VA gap junction, suggesting that the best circuit's use of it may have been unusual. Otherwise, the remaining connections also made only minor contributions to the solutions in the ensemble: VA$\to$VD was unnecessary in 75\% of the solutions, DA$\to$VD was unnecessary in 76\% of the solutions, DB$\to$VD was unnecessary in 99\% of the solutions, and the  VD$\vdashv$VD and VB$\vdashv$VB gap junctions were always unnecessary (Fig~\ref{fig_EnsembleAblation}C). 

\subsection*{Minimal functional subcircuit present in connectome}

\begin{figure}[!ht]
% \widefigure{\fullpagewidth}{figures/Fig11} 
\includegraphics[width=13.2cm]{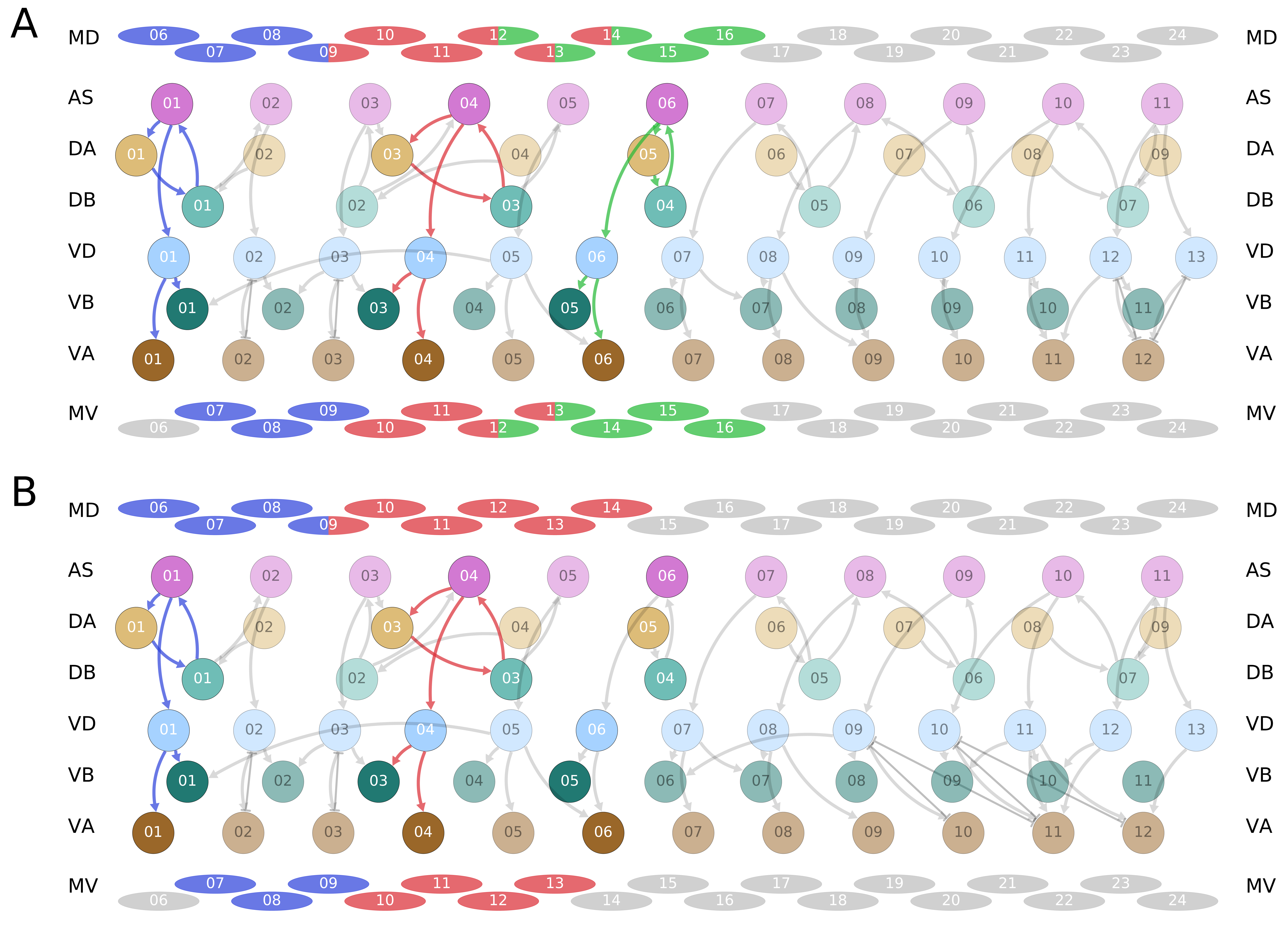}
\centering
\caption{{\bf Minimal functional subcircuit present in connectome.}
The circuit identified in the analysis exists in three places in the anterior section of the connectome of the \textit{C.~elegans} hermaphrodite [A] and twice in the anterior section of the male connectome [B]. Each functional subcircuit is shown with a different color (e.g., blue, red, and green). The muscles innervated by these subcircuits are colored accordingly. Neurons and muscles are depicted along the VNC according to their sequential order (not exact body position). Light shaded connections correspond to synapses that were identified as important for the minimal functional subcircuit along the rest of the VNC but that do not form a complete subcircuit. For clarity, all other chemical synapses and gap junctions are not shown.} 
\label{fig_CPGFullConnectome}
\end{figure}

It is important to recall that the repeating neural unit upon which we based our model (Fig~\ref{fig_HaspelSegment}) is only a statistical summary of the VNC~\citep{Haspel2011}. How well do the key components that we have identified map onto the actual neuroanatomy? In order to answer this question, we examined the most recent reconstructions of both the hermaphrodite and male nematodes~\citep{Jarrell2012,Xu2013} for the existence of the two forms of the minimal functional subcircuit found in the ensemble (Fig~\ref{fig_TwoMinimalFunctionalCircuits}). We found that all of the key components of one of the forms of the minimal functional subcircuit occur together in three different places in the hermaphrodite (Fig~\ref{fig_CPGFullConnectome}A) and twice in the male connectome (Fig~\ref{fig_CPGFullConnectome}B). This includes the AS-DA-DB dorsal core oscillator, the AS$\to$VD connection responsible for antiphase dorsoventral propagation, and the VD$\to$VA and VD$\to$VB connections that propagate the oscillation to the remaining ventral motorneurons. From this result, we make three main observations. First, although the most anterior instance aligns perfectly with the perimotor segmentation proposed by~\citep{Haspel2011}, the other two do not: they include neurons across adjacent units of the VNC. This suggests CPGs might be occurring across the repeating neural units originally proposed. Second, the motorneurons in each of the three instances of the minimal functional circuit innervate muscles that are relatively close to each other. Such an alignment could provide a solid foundation for driving movement within a significant segment of the body. Finally, altogether the three subcircuits innervate muscles spanning the full anterior body (Fig~\ref{fig_CPGFullConnectome}). This makes sense given the important role that stretch receptors have been shown to play in propagating the wave posteriorly during forward locomotion~\citep{Wen2012}.

%%%%%%%%%%%%  Figure minimal network found in connectome     %%%%%%%%%%%%

%%%%%%%%%%%%%%%%%%%%%%%%%%%%%%%%%%%%%%%%%%%%%%%%%%%%%%%%%%%%
%%%%%%%%%%%%%%%%%%%%%%%%%%%%%%%%%%%%%%%%%%%%%%%%%%%%%%%%%%%%
%%%%%%%%%%%%%%%%%%%%%%%%%%%%%%%%%%%%%%%%%%%%%%%%%%%%%%%%%%%%

\section*{Discussion}
The goal of this work was to test whether the VNC is capable of intrinsically-generating oscillations that match what has been observed in A- and B-class motorneurons during forward and backward locomotion~\citep{Haspel2010,Kawano2011,Faumont2011}. In order to address this, we applied evolutionary algorithms to systematically search for circuits that could generate intrinsic network oscillations within the worm's ventral nerve cord. We demonstrate for the first time that CPGs that match some of the most salient features of forward and backward neural traces are possible in the VNC, given what we know about the neuroanatomy and neurophysiology of \textit{C.~elegans}. Although we have made a number of simplifying assumptions in order to develop a neuroanatomically-grounded computational model of the ventral nerve cord, these do not undermine our main result: that a VNC CPG for locomotion is feasible. Our modeling approach was deliberately minimal given our goal. Adding omitted chemical synapses, gap junctions, and motorneurons, removing the anterior-posterior symmetry constraint on the connections in the neural unit, and including additional electrophysiological details from the  motorneurons and command interneurons has the potential to enhance the richness of available mechanisms for reproducing matching oscillatory activity. Adding omitted chemical synapses, gap junctions, and motorneurons, removing the anterior-posterior symmetry constraint on the connections in the neural unit, and including additional electrophysiological details from the motorneurons and command interneurons has the potential to enhance the richness of available mechanisms for reproducing matching oscillatory activity.

Analysis of the best evolved circuit revealed a dorsal core subcircuit composed of an AS, a DA, and a DB motorneuron, altogether responsible for the generation of the oscillation. Through the use of our optimization methodology, we also revealed a number of different sign patterns in which the dorsal core oscillator could be instantiated. Although much less common, there were two other kinds of solutions found in the ensemble that were different from dorsal core subcircuits: solutions that generated oscillations through a ventral core comprising VA and VD; and a solution that required all motorneurons except AS and DD to produce oscillations. Our analysis also revealed a minimal functional circuit which included a VD, a VA, and a VB motorneuron in addition to the dorsal core, which exhibited all the main features observed in the worm's neural traces. Interestingly, analysis of the ensemble of successful solutions revealed that the minimal functional circuit was one of the most common solutions. Based on our findings from the model, we returned to the connectome reconstructions and found multiple instances of this subcircuit in the hermaphrodite and male. 

Given the existence of subcircuits in the VNC capable of producing backward-forward oscillatory activity, our model makes a set of testable predictions for the worm. Short of being able to suppress stretch receptor activity during locomotion, to see the extent to which the worm can move forward, one of the key experiments that emerge from our analysis is to selectively suppress activity in dorsal-only or ventral-only motorneurons. Based on the role of the dorsal core subcircuit, our model suggests that suppressing only ventral motorneurons would have a smaller effect than suppressing only dorsal motorneurons. A second key experimental effort involves characterizing the physiology of the AS motorneuron. If a dorsal core is present in the worm, then our model suggests AS is likely to be a bistable motorneuron~\citep{Mellem2008}, but not DA or DB. A third key experiment involves disrupting the chemical synapse from AS to VD. Regardless of how oscillations are generated, if the pattern of activity is present in the dorsal motorneurons, our model suggests it can be spread to ventral motorneurons through the connection between AS and VD to produce dorsoventral antiphase patterns of activity. Finally, based on our analysis, network oscillators are most feasible in the anterior portion of the VNC. Although further investigations would be required, our evolutionary experiments did not identify a CPG subcircuit in the posterior half of the VNC. In the theory-experiment cycle, results from additional experiments in the worm will help add additional constraints to our model. 

Three hypotheses for the generation of oscillations for locomotion in \textit{C. elegans} have been postulated~\citep{Gjorgjieva2014,Cohen2014,Zhen2015}: (a) Via stretch-receptor feedback; (b) Via a single CPG in the head circuit; (c) Via one or more CPGs along the ventral nerve cord. Until now, only the first of these two hypotheses had been demonstrated using neuroanatomically-grounded computational models~\citep{Bryden2008,Karbowski2008,Niebur1993,Niebur1991,Sakata2005,Wen2012,Deng2016,Boyle2008,Boyle2012,Kunert2017}. By demonstrating that intrinsic network oscillations in the ventral nerve cord are possible, we are elevating the previously neglected VNC CPG hypothesis to the level of feasibility of the other two more commonly discussed locomotion hypotheses. Ultimately, it is important to keep in mind that these and other related hypotheses are not mutually exclusive: the existence of a network oscillator in the VNC is compatible with many of the other mechanisms that have been suggested to operate during locomotion. First, a CPG in the worm's head could entrain the CPGs in the ventral nerve cord. Recent observations of different oscillatory rhythms in the head and the rest of the body suggests this is indeed a possibility~\citep{Fouad2017}. Although we found one set of motifs that work as a CPG and is present in the anterior half of the worm, it is possible that there are other CPG motifs or reflexive pattern generators that exist in the rest of the body, including some that we have not found yet. Second, stretch-receptor feedback~\citep{Wen2012} could modulate intrinsic network oscillations from CPGs in the ventral nerve cord, effectively generating different oscillating frequencies when the worm is placed in different media~\citep{Berri2009,Fang2010,Lebois2012}. Third, signals from upstream command interneurons and the inter-unit electrical coupling within the VNC could help coordinate multiple CPGs in the VNC and help modulate their intrinsic network oscillations~\citep{Xu2017}. Finally, if pacemaker neurons exist in the ventral nerve cord, they could aide the maintenance of intrinsic network oscillations within a neural unit~\citep{Gao2017}.

Ultimately, understanding how neural circuits produce behavior requires an understanding of the complete brain-body-environment system~\citep{Chiel1997}. Although we have modeled the worm's body in previous contributions~\citep{Izquierdo2015}, in the current model, only motorneurons were modeled. In the worm, these motorneurons are connected to the muscles through known neuromuscular junctions, which in turn drive the worm mechanically to produce thrust against the surface of its immediate environment. For any of the solutions revealed in our study to drive locomotion, the strengths of the neuromuscular junction would have to be tuned to effectively modulate the muscles.  Furthermore, for the wave to travel posteriorly along the VNC, the subcircuits would have to coordinate with each other. This could be done via chemical synapses and gap junctions within the nervous system, or through the use of stretch receptors, via the mechanical movement of the body, or a combination of both. Whether forward and backward CPG-driven locomotion is feasible is unknown, but we are now in a position to test these ideas by integrating these circuits in a neuromechanical model of locomotion in the worm.

\section*{Acknowledgments}
The work in this paper was supported in part by NSF grant IIS-1524647.

\bibliography{references}

\begin{thebibliography}{}

\bibitem [\protect \citeauthoryear {%
Beer%
}{%
Beer%
}{%
{\protect \APACyear {1995}}%
}]{%
Beer1995}
\APACinsertmetastar {%
Beer1995}%
\begin{APACrefauthors}%
Beer, R\BPBI D.%
\end{APACrefauthors}%
\unskip\
\newblock
\APACrefYearMonthDay{1995}{}{}.
\newblock
{\BBOQ}\APACrefatitle {On the dynamics of small continuous-time recurrent
  neural networks} {On the dynamics of small continuous-time recurrent neural
  networks}.{\BBCQ}
\newblock
\APACjournalVolNumPages{Adapt Behav}{3}{4}{469--509}.
\PrintBackRefs{\CurrentBib}

\bibitem [\protect \citeauthoryear {%
Berri%
, Boyle%
, Tassieri%
, Hope%
\BCBL {}\ \BBA {} Cohen%
}{%
Berri%
\ \protect \BOthers {.}}{%
{\protect \APACyear {2009}}%
}]{%
Berri2009}
\APACinsertmetastar {%
Berri2009}%
\begin{APACrefauthors}%
Berri, S.%
, Boyle, J\BPBI H.%
, Tassieri, M.%
, Hope, I\BPBI A.%
\BCBL {}\ \BBA {} Cohen, N.%
\end{APACrefauthors}%
\unskip\
\newblock
\APACrefYearMonthDay{2009}{}{}.
\newblock
{\BBOQ}\APACrefatitle {Forward locomotion of the nematode \textit{C. elegans}
  is achieved through modulation of a single gait} {Forward locomotion of the
  nematode \textit{C. elegans} is achieved through modulation of a single
  gait}.{\BBCQ}
\newblock
\APACjournalVolNumPages{HFSP J}{3}{3}{186--193}.
\PrintBackRefs{\CurrentBib}

\bibitem [\protect \citeauthoryear {%
Boyle%
, Berri%
\BCBL {}\ \BBA {} Cohen%
}{%
Boyle%
\ \protect \BOthers {.}}{%
{\protect \APACyear {2012}}%
}]{%
Boyle2012}
\APACinsertmetastar {%
Boyle2012}%
\begin{APACrefauthors}%
Boyle, J\BPBI H.%
, Berri, S.%
\BCBL {}\ \BBA {} Cohen, N.%
\end{APACrefauthors}%
\unskip\
\newblock
\APACrefYearMonthDay{2012}{}{}.
\newblock
{\BBOQ}\APACrefatitle {Gait modulation in \textit{C. elegans}: an integrated
  neuromechanical model} {Gait modulation in \textit{C. elegans}: an integrated
  neuromechanical model}.{\BBCQ}
\newblock
\APACjournalVolNumPages{Front Comput Neurosci}{6}{}{10}.
\newblock
\begin{APACrefDOI} \doi{10.3389/fncom.2012.00010} \end{APACrefDOI}
\PrintBackRefs{\CurrentBib}

\bibitem [\protect \citeauthoryear {%
Boyle%
, Bryden%
\BCBL {}\ \BBA {} Cohen%
}{%
Boyle%
\ \protect \BOthers {.}}{%
{\protect \APACyear {2008}}%
}]{%
Boyle2008}
\APACinsertmetastar {%
Boyle2008}%
\begin{APACrefauthors}%
Boyle, J\BPBI H.%
, Bryden, J.%
\BCBL {}\ \BBA {} Cohen, N.%
\end{APACrefauthors}%
\unskip\
\newblock
\APACrefYearMonthDay{2008}{}{}.
\newblock
{\BBOQ}\APACrefatitle {An integrated neuro-mechanical model of \textit{C.
  elegans} forward locomotion} {An integrated neuro-mechanical model of
  \textit{C. elegans} forward locomotion}.{\BBCQ}
\newblock
\BIn{} M.~Ishikawa, K.~Doya, H.~Miyamoto\BCBL {}\ \BBA {} T.~Yamakawa\ (\BEDS),
  \APACrefbtitle {Neural Information Processing. ICONIP 2007. Lecture Notes in
  Computer Science} {Neural information processing. iconip 2007. lecture notes
  in computer science}\ (\BPGS\ 37--47).
\newblock
\APACaddressPublisher{Berlin, Heidelberg}{Springer}.
\PrintBackRefs{\CurrentBib}

\bibitem [\protect \citeauthoryear {%
Bryden%
\ \BBA {} Cohen%
}{%
Bryden%
\ \BBA {} Cohen%
}{%
{\protect \APACyear {2008}}%
}]{%
Bryden2008}
\APACinsertmetastar {%
Bryden2008}%
\begin{APACrefauthors}%
Bryden, J.%
\BCBT {}\ \BBA {} Cohen, N.%
\end{APACrefauthors}%
\unskip\
\newblock
\APACrefYearMonthDay{2008}{}{}.
\newblock
{\BBOQ}\APACrefatitle {Neural control of \textit{Caenorhabditis elegans}
  forward locomotion: the role of sensory feedback} {Neural control of
  \textit{Caenorhabditis elegans} forward locomotion: the role of sensory
  feedback}.{\BBCQ}
\newblock
\APACjournalVolNumPages{Biol Cybern}{98}{4}{339--351}.
\PrintBackRefs{\CurrentBib}

\bibitem [\protect \citeauthoryear {%
Chalfie%
\ \protect \BOthers {.}}{%
Chalfie%
\ \protect \BOthers {.}}{%
{\protect \APACyear {1985}}%
}]{%
Chalfie1985}
\APACinsertmetastar {%
Chalfie1985}%
\begin{APACrefauthors}%
Chalfie, M.%
, Sulston, J\BPBI E.%
, White, J\BPBI G.%
, Southgate, E.%
, Thomson, J\BPBI N.%
\BCBL {}\ \BBA {} Brenner, S.%
\end{APACrefauthors}%
\unskip\
\newblock
\APACrefYearMonthDay{1985}{}{}.
\newblock
{\BBOQ}\APACrefatitle {The neural circuit for touch sensitivity in
  \textit{Caenorhabditis elegans}} {The neural circuit for touch sensitivity in
  \textit{Caenorhabditis elegans}}.{\BBCQ}
\newblock
\APACjournalVolNumPages{J Neurosci}{5}{4}{956--964}.
\PrintBackRefs{\CurrentBib}

\bibitem [\protect \citeauthoryear {%
Chen%
, Hall%
\BCBL {}\ \BBA {} Chklovskii%
}{%
Chen%
\ \protect \BOthers {.}}{%
{\protect \APACyear {2006}}%
}]{%
Chen2006}
\APACinsertmetastar {%
Chen2006}%
\begin{APACrefauthors}%
Chen, B.%
, Hall, D.%
\BCBL {}\ \BBA {} Chklovskii, D.%
\end{APACrefauthors}%
\unskip\
\newblock
\APACrefYearMonthDay{2006}{}{}.
\newblock
{\BBOQ}\APACrefatitle {Wiring optimization can relate neuronal structure and
  function} {Wiring optimization can relate neuronal structure and
  function}.{\BBCQ}
\newblock
\APACjournalVolNumPages{Proc Natl Acad Sci USA}{103}{12}{4723--4728}.
\PrintBackRefs{\CurrentBib}

\bibitem [\protect \citeauthoryear {%
Chiel%
\ \BBA {} Beer%
}{%
Chiel%
\ \BBA {} Beer%
}{%
{\protect \APACyear {1997}}%
}]{%
Chiel1997}
\APACinsertmetastar {%
Chiel1997}%
\begin{APACrefauthors}%
Chiel, H\BPBI J.%
\BCBT {}\ \BBA {} Beer, R\BPBI D.%
\end{APACrefauthors}%
\unskip\
\newblock
\APACrefYearMonthDay{1997}{}{}.
\newblock
{\BBOQ}\APACrefatitle {The brain has a body: adaptive behavior emerges from
  interactions of nervous system, body and environment} {The brain has a body:
  adaptive behavior emerges from interactions of nervous system, body and
  environment}.{\BBCQ}
\newblock
\APACjournalVolNumPages{Trends Neurosci}{20}{12}{553--557}.
\PrintBackRefs{\CurrentBib}

\bibitem [\protect \citeauthoryear {%
Cohen%
\ \BBA {} Sanders%
}{%
Cohen%
\ \BBA {} Sanders%
}{%
{\protect \APACyear {2014}}%
}]{%
Cohen2014}
\APACinsertmetastar {%
Cohen2014}%
\begin{APACrefauthors}%
Cohen, N.%
\BCBT {}\ \BBA {} Sanders, T.%
\end{APACrefauthors}%
\unskip\
\newblock
\APACrefYearMonthDay{2014}{}{}.
\newblock
{\BBOQ}\APACrefatitle {Nematode locomotion: dissecting the
  neuronal--environmental loop} {Nematode locomotion: dissecting the
  neuronal--environmental loop}.{\BBCQ}
\newblock
\APACjournalVolNumPages{Curr Opin Neurobiol}{25}{}{99--106}.
\PrintBackRefs{\CurrentBib}

\bibitem [\protect \citeauthoryear {%
Deng%
, Xu%
, Wang%
, Wang%
\BCBL {}\ \BBA {} Chen%
}{%
Deng%
\ \protect \BOthers {.}}{%
{\protect \APACyear {2016}}%
}]{%
Deng2016}
\APACinsertmetastar {%
Deng2016}%
\begin{APACrefauthors}%
Deng, X.%
, Xu, J\BHBI X.%
, Wang, J.%
, Wang, G\BHBI y.%
\BCBL {}\ \BBA {} Chen, Q\BHBI s.%
\end{APACrefauthors}%
\unskip\
\newblock
\APACrefYearMonthDay{2016}{}{}.
\newblock
{\BBOQ}\APACrefatitle {Biological modeling the undulatory locomotion of
  \textit{C. elegans} using dynamic neural network approach} {Biological
  modeling the undulatory locomotion of \textit{C. elegans} using dynamic
  neural network approach}.{\BBCQ}
\newblock
\APACjournalVolNumPages{Neurocomputing}{186}{}{207--217}.
\PrintBackRefs{\CurrentBib}

\bibitem [\protect \citeauthoryear {%
Dimitrijevic%
, Gerasimenko%
\BCBL {}\ \BBA {} Pinter%
}{%
Dimitrijevic%
\ \protect \BOthers {.}}{%
{\protect \APACyear {1998}}%
}]{%
Dimitrijevic1998}
\APACinsertmetastar {%
Dimitrijevic1998}%
\begin{APACrefauthors}%
Dimitrijevic, M\BPBI R.%
, Gerasimenko, Y.%
\BCBL {}\ \BBA {} Pinter, M\BPBI M.%
\end{APACrefauthors}%
\unskip\
\newblock
\APACrefYearMonthDay{1998}{}{}.
\newblock
{\BBOQ}\APACrefatitle {Evidence for a spinal central pattern generator in
  humans} {Evidence for a spinal central pattern generator in humans}.{\BBCQ}
\newblock
\APACjournalVolNumPages{Ann N Y Acad Sci}{860}{1}{360--376}.
\PrintBackRefs{\CurrentBib}

\bibitem [\protect \citeauthoryear {%
Fang-Yen%
\ \protect \BOthers {.}}{%
Fang-Yen%
\ \protect \BOthers {.}}{%
{\protect \APACyear {2010}}%
}]{%
Fang2010}
\APACinsertmetastar {%
Fang2010}%
\begin{APACrefauthors}%
Fang-Yen, C.%
, Wyart, M.%
, Xie, J.%
, Kawai, R.%
, Kodger, T.%
, Chen, S.%
\BDBL {}Samuel, A\BPBI D.%
\end{APACrefauthors}%
\unskip\
\newblock
\APACrefYearMonthDay{2010}{}{}.
\newblock
{\BBOQ}\APACrefatitle {Biomechanical analysis of gait adaptation in the
  nematode \textit{Caenorhabditis elegans}} {Biomechanical analysis of gait
  adaptation in the nematode \textit{Caenorhabditis elegans}}.{\BBCQ}
\newblock
\APACjournalVolNumPages{Proc Natl Acad Sci USA}{107}{47}{20323--20328}.
\PrintBackRefs{\CurrentBib}

\bibitem [\protect \citeauthoryear {%
Faumont%
\ \protect \BOthers {.}}{%
Faumont%
\ \protect \BOthers {.}}{%
{\protect \APACyear {2011}}%
}]{%
Faumont2011}
\APACinsertmetastar {%
Faumont2011}%
\begin{APACrefauthors}%
Faumont, S.%
, Rondeau, G.%
, Thiele, T\BPBI R.%
, Lawton, K\BPBI J.%
, McCormick, K\BPBI E.%
, Sottile, M.%
\BDBL {}others%
\end{APACrefauthors}%
\unskip\
\newblock
\APACrefYearMonthDay{2011}{}{}.
\newblock
{\BBOQ}\APACrefatitle {An image-free opto-mechanical system for creating
  virtual environments and imaging neuronal activity in freely moving
  \textit{Caenorhabditis elegans}} {An image-free opto-mechanical system for
  creating virtual environments and imaging neuronal activity in freely moving
  \textit{Caenorhabditis elegans}}.{\BBCQ}
\newblock
\APACjournalVolNumPages{PLoS One}{6}{9}{e24666}.
\newblock
\begin{APACrefDOI} \doi{10.1371/journal.pone.0024666} \end{APACrefDOI}
\PrintBackRefs{\CurrentBib}

\bibitem [\protect \citeauthoryear {%
Fouad%
\ \protect \BOthers {.}}{%
Fouad%
\ \protect \BOthers {.}}{%
{\protect \APACyear {2017}}%
}]{%
Fouad2017}
\APACinsertmetastar {%
Fouad2017}%
\begin{APACrefauthors}%
Fouad, A\BPBI D.%
, Teng, S.%
, Mark, J\BPBI R.%
, Liu, A.%
, Ji, H.%
, Cornblath, E.%
\BDBL {}Fang-Yen, C.%
\end{APACrefauthors}%
\unskip\
\newblock
\APACrefYearMonthDay{2017}{}{}.
\newblock
{\BBOQ}\APACrefatitle {Distributed rhythm generators underlie
  \textit{Caenorhabditis elegans} forward locomotion} {Distributed rhythm
  generators underlie \textit{Caenorhabditis elegans} forward
  locomotion}.{\BBCQ}
\newblock
\APACjournalVolNumPages{bioRxiv}{}{}{}.
\newblock
\begin{APACrefDOI} \doi{10.1101/141911} \end{APACrefDOI}
\PrintBackRefs{\CurrentBib}

\bibitem [\protect \citeauthoryear {%
Gao%
\ \protect \BOthers {.}}{%
Gao%
\ \protect \BOthers {.}}{%
{\protect \APACyear {2017}}%
}]{%
Gao2017}
\APACinsertmetastar {%
Gao2017}%
\begin{APACrefauthors}%
Gao, S.%
, Guan, S\BPBI A.%
, Fouad, A\BPBI D.%
, Meng, J.%
, Huang, Y\BHBI C.%
, Li, Y.%
\BDBL {}Zhen, M.%
\end{APACrefauthors}%
\unskip\
\newblock
\APACrefYearMonthDay{2017}{}{}.
\newblock
{\BBOQ}\APACrefatitle {Excitatory motor neurons are local central pattern
  generators in an anatomically compressed motor circuit for reverse
  locomotion} {Excitatory motor neurons are local central pattern generators in
  an anatomically compressed motor circuit for reverse locomotion}.{\BBCQ}
\newblock
\APACjournalVolNumPages{bioRxiv}{}{}{}.
\newblock
\begin{APACrefDOI} \doi{10.1101/135418} \end{APACrefDOI}
\PrintBackRefs{\CurrentBib}

\bibitem [\protect \citeauthoryear {%
Gjorgjieva%
, Biron%
\BCBL {}\ \BBA {} Haspel%
}{%
Gjorgjieva%
\ \protect \BOthers {.}}{%
{\protect \APACyear {2014}}%
}]{%
Gjorgjieva2014}
\APACinsertmetastar {%
Gjorgjieva2014}%
\begin{APACrefauthors}%
Gjorgjieva, J.%
, Biron, D.%
\BCBL {}\ \BBA {} Haspel, G.%
\end{APACrefauthors}%
\unskip\
\newblock
\APACrefYearMonthDay{2014}{}{}.
\newblock
{\BBOQ}\APACrefatitle {Neurobiology of \textit{Caenorhabditis elegans}
  locomotion: where do we stand?} {Neurobiology of \textit{Caenorhabditis
  elegans} locomotion: where do we stand?}{\BBCQ}
\newblock
\APACjournalVolNumPages{Bioscience}{64}{6}{476--486}.
\PrintBackRefs{\CurrentBib}

\bibitem [\protect \citeauthoryear {%
Gray%
\ \BBA {} Lissmann%
}{%
Gray%
\ \BBA {} Lissmann%
}{%
{\protect \APACyear {1964}}%
}]{%
Gray1964}
\APACinsertmetastar {%
Gray1964}%
\begin{APACrefauthors}%
Gray, J.%
\BCBT {}\ \BBA {} Lissmann, H.%
\end{APACrefauthors}%
\unskip\
\newblock
\APACrefYearMonthDay{1964}{}{}.
\newblock
{\BBOQ}\APACrefatitle {The locomotion of nematodes} {The locomotion of
  nematodes}.{\BBCQ}
\newblock
\APACjournalVolNumPages{J Exp Biol}{41}{}{135--154}.
\PrintBackRefs{\CurrentBib}

\bibitem [\protect \citeauthoryear {%
Guertin%
}{%
Guertin%
}{%
{\protect \APACyear {2013}}%
}]{%
Guertin2013}
\APACinsertmetastar {%
Guertin2013}%
\begin{APACrefauthors}%
Guertin, P\BPBI A.%
\end{APACrefauthors}%
\unskip\
\newblock
\APACrefYearMonthDay{2013}{}{}.
\newblock
{\BBOQ}\APACrefatitle {Central pattern generator for locomotion: anatomical,
  physiological, and pathophysiological considerations} {Central pattern
  generator for locomotion: anatomical, physiological, and pathophysiological
  considerations}.{\BBCQ}
\newblock
\APACjournalVolNumPages{Front Neurol}{3}{}{183}.
\newblock
\begin{APACrefDOI} \doi{10.3389/fneur.2012.00183} \end{APACrefDOI}
\PrintBackRefs{\CurrentBib}

\bibitem [\protect \citeauthoryear {%
Haspel%
\ \BBA {} O'Donovan%
}{%
Haspel%
\ \BBA {} O'Donovan%
}{%
{\protect \APACyear {2011}}%
}]{%
Haspel2011}
\APACinsertmetastar {%
Haspel2011}%
\begin{APACrefauthors}%
Haspel, G.%
\BCBT {}\ \BBA {} O'Donovan, M\BPBI J.%
\end{APACrefauthors}%
\unskip\
\newblock
\APACrefYearMonthDay{2011}{}{}.
\newblock
{\BBOQ}\APACrefatitle {A perimotor framework reveals functional segmentation in
  the motoneuronal network controlling locomotion in \textit{Caenorhabditis
  elegans}} {A perimotor framework reveals functional segmentation in the
  motoneuronal network controlling locomotion in \textit{Caenorhabditis
  elegans}}.{\BBCQ}
\newblock
\APACjournalVolNumPages{J Neurosci}{31}{41}{14611--14623}.
\PrintBackRefs{\CurrentBib}

\bibitem [\protect \citeauthoryear {%
Haspel%
, O'Donovan%
\BCBL {}\ \BBA {} Hart%
}{%
Haspel%
\ \protect \BOthers {.}}{%
{\protect \APACyear {2010}}%
}]{%
Haspel2010}
\APACinsertmetastar {%
Haspel2010}%
\begin{APACrefauthors}%
Haspel, G.%
, O'Donovan, M\BPBI J.%
\BCBL {}\ \BBA {} Hart, A\BPBI C.%
\end{APACrefauthors}%
\unskip\
\newblock
\APACrefYearMonthDay{2010}{}{}.
\newblock
{\BBOQ}\APACrefatitle {Motoneurons dedicated to either forward or backward
  locomotion in the nematode \textit{Caenorhabditis elegans}} {Motoneurons
  dedicated to either forward or backward locomotion in the nematode
  \textit{Caenorhabditis elegans}}.{\BBCQ}
\newblock
\APACjournalVolNumPages{J Neurosci}{30}{33}{11151--11156}.
\PrintBackRefs{\CurrentBib}

\bibitem [\protect \citeauthoryear {%
Ijspeert%
}{%
Ijspeert%
}{%
{\protect \APACyear {2008}}%
}]{%
Ijspeert2008}
\APACinsertmetastar {%
Ijspeert2008}%
\begin{APACrefauthors}%
Ijspeert, A\BPBI J.%
\end{APACrefauthors}%
\unskip\
\newblock
\APACrefYearMonthDay{2008}{}{}.
\newblock
{\BBOQ}\APACrefatitle {Central pattern generators for locomotion control in
  animals and robots: a review} {Central pattern generators for locomotion
  control in animals and robots: a review}.{\BBCQ}
\newblock
\APACjournalVolNumPages{Neural Netw}{21}{4}{642--653}.
\PrintBackRefs{\CurrentBib}

\bibitem [\protect \citeauthoryear {%
Izquierdo%
\ \BBA {} Beer%
}{%
Izquierdo%
\ \BBA {} Beer%
}{%
{\protect \APACyear {2013}}%
}]{%
Izquierdo2013}
\APACinsertmetastar {%
Izquierdo2013}%
\begin{APACrefauthors}%
Izquierdo, E.%
\BCBT {}\ \BBA {} Beer, R.%
\end{APACrefauthors}%
\unskip\
\newblock
\APACrefYearMonthDay{2013}{}{}.
\newblock
{\BBOQ}\APACrefatitle {Connecting a connectome to behavior: an ensemble of
  neuroanatomical models of \textit{C. elegans} klinotaxis} {Connecting a
  connectome to behavior: an ensemble of neuroanatomical models of \textit{C.
  elegans} klinotaxis}.{\BBCQ}
\newblock
\APACjournalVolNumPages{PLoS Comput Biol}{9}{2}{e1002890}.
\newblock
\begin{APACrefDOI} \doi{10.1371/journal.pcbi.1002890} \end{APACrefDOI}
\PrintBackRefs{\CurrentBib}

\bibitem [\protect \citeauthoryear {%
Izquierdo%
\ \BBA {} Beer%
}{%
Izquierdo%
\ \BBA {} Beer%
}{%
{\protect \APACyear {2015}}%
}]{%
Izquierdo2015}
\APACinsertmetastar {%
Izquierdo2015}%
\begin{APACrefauthors}%
Izquierdo, E.%
\BCBT {}\ \BBA {} Beer, R.%
\end{APACrefauthors}%
\unskip\
\newblock
\APACrefYearMonthDay{2015}{}{}.
\newblock
{\BBOQ}\APACrefatitle {An integrated neuromechanical model of steering in
  \textit{C elegans}} {An integrated neuromechanical model of steering in
  \textit{C elegans}}.{\BBCQ}
\newblock
\BIn{} P.~Andrews\ \BOthers {.}\ (\BEDS), \APACrefbtitle {Proceedings of the
  European Conference on Artificial Life 2015, ECAL 2015} {Proceedings of the
  european conference on artificial life 2015, ecal 2015}\ (\BPGS\ 199--206).
\newblock
\APACaddressPublisher{}{MIT Press}.
\PrintBackRefs{\CurrentBib}

\bibitem [\protect \citeauthoryear {%
Izquierdo%
\ \BBA {} Lockery%
}{%
Izquierdo%
\ \BBA {} Lockery%
}{%
{\protect \APACyear {2010}}%
}]{%
Izquierdo2010}
\APACinsertmetastar {%
Izquierdo2010}%
\begin{APACrefauthors}%
Izquierdo, E.%
\BCBT {}\ \BBA {} Lockery, S.%
\end{APACrefauthors}%
\unskip\
\newblock
\APACrefYearMonthDay{2010}{}{}.
\newblock
{\BBOQ}\APACrefatitle {Evolution and analysis of minimal neural circuits for
  klinotaxis in \textit{Caenorhabditis elegans}} {Evolution and analysis of
  minimal neural circuits for klinotaxis in \textit{Caenorhabditis
  elegans}}.{\BBCQ}
\newblock
\APACjournalVolNumPages{J Neurosci}{30}{39}{12908--12917}.
\PrintBackRefs{\CurrentBib}

\bibitem [\protect \citeauthoryear {%
Jarrell%
\ \protect \BOthers {.}}{%
Jarrell%
\ \protect \BOthers {.}}{%
{\protect \APACyear {2012}}%
}]{%
Jarrell2012}
\APACinsertmetastar {%
Jarrell2012}%
\begin{APACrefauthors}%
Jarrell, T\BPBI A.%
, Wang, Y.%
, Bloniarz, A\BPBI E.%
, Brittin, C\BPBI A.%
, Xu, M.%
, Thomson, J\BPBI N.%
\BDBL {}Emmons, S\BPBI W.%
\end{APACrefauthors}%
\unskip\
\newblock
\APACrefYearMonthDay{2012}{}{}.
\newblock
{\BBOQ}\APACrefatitle {The connectome of a decision-making neural network} {The
  connectome of a decision-making neural network}.{\BBCQ}
\newblock
\APACjournalVolNumPages{Science}{337}{6093}{437--444}.
\PrintBackRefs{\CurrentBib}

\bibitem [\protect \citeauthoryear {%
Karbowski%
, Schindelman%
, Cronin%
, Seah%
\BCBL {}\ \BBA {} Sternberg%
}{%
Karbowski%
\ \protect \BOthers {.}}{%
{\protect \APACyear {2008}}%
}]{%
Karbowski2008}
\APACinsertmetastar {%
Karbowski2008}%
\begin{APACrefauthors}%
Karbowski, J.%
, Schindelman, G.%
, Cronin, C\BPBI J.%
, Seah, A.%
\BCBL {}\ \BBA {} Sternberg, P\BPBI W.%
\end{APACrefauthors}%
\unskip\
\newblock
\APACrefYearMonthDay{2008}{}{}.
\newblock
{\BBOQ}\APACrefatitle {Systems level circuit model of \textit{C. elegans}
  undulatory locomotion: mathematical modeling and molecular genetics} {Systems
  level circuit model of \textit{C. elegans} undulatory locomotion:
  mathematical modeling and molecular genetics}.{\BBCQ}
\newblock
\APACjournalVolNumPages{J Comput Neurosci}{24}{3}{253--276}.
\PrintBackRefs{\CurrentBib}

\bibitem [\protect \citeauthoryear {%
Kato%
\ \protect \BOthers {.}}{%
Kato%
\ \protect \BOthers {.}}{%
{\protect \APACyear {2015}}%
}]{%
kato2015global}
\APACinsertmetastar {%
kato2015global}%
\begin{APACrefauthors}%
Kato, S.%
, Kaplan, H\BPBI S.%
, Schr{\"o}del, T.%
, Skora, S.%
, Lindsay, T\BPBI H.%
, Yemini, E.%
\BDBL {}Zimmer, M.%
\end{APACrefauthors}%
\unskip\
\newblock
\APACrefYearMonthDay{2015}{}{}.
\newblock
{\BBOQ}\APACrefatitle {Global brain dynamics embed the motor command sequence
  of Caenorhabditis elegans} {Global brain dynamics embed the motor command
  sequence of caenorhabditis elegans}.{\BBCQ}
\newblock
\APACjournalVolNumPages{Cell}{163}{3}{656--669}.
\PrintBackRefs{\CurrentBib}

\bibitem [\protect \citeauthoryear {%
Kawano%
\ \protect \BOthers {.}}{%
Kawano%
\ \protect \BOthers {.}}{%
{\protect \APACyear {2011}}%
}]{%
Kawano2011}
\APACinsertmetastar {%
Kawano2011}%
\begin{APACrefauthors}%
Kawano, T.%
, Po, M\BPBI D.%
, Gao, S.%
, Leung, G.%
, Ryu, W\BPBI S.%
\BCBL {}\ \BBA {} Zhen, M.%
\end{APACrefauthors}%
\unskip\
\newblock
\APACrefYearMonthDay{2011}{}{}.
\newblock
{\BBOQ}\APACrefatitle {An imbalancing act: gap junctions reduce the backward
  motor circuit activity to bias \textit{C. elegans} for forward locomotion}
  {An imbalancing act: gap junctions reduce the backward motor circuit activity
  to bias \textit{C. elegans} for forward locomotion}.{\BBCQ}
\newblock
\APACjournalVolNumPages{Neuron}{72}{4}{572--586}.
\PrintBackRefs{\CurrentBib}

\bibitem [\protect \citeauthoryear {%
Kunert%
, Proctor%
, Brunton%
\BCBL {}\ \BBA {} Kutz%
}{%
Kunert%
\ \protect \BOthers {.}}{%
{\protect \APACyear {2017}}%
}]{%
Kunert2017}
\APACinsertmetastar {%
Kunert2017}%
\begin{APACrefauthors}%
Kunert, J.%
, Proctor, J.%
, Brunton, S.%
\BCBL {}\ \BBA {} Kutz, J.%
\end{APACrefauthors}%
\unskip\
\newblock
\APACrefYearMonthDay{2017}{}{}.
\newblock
{\BBOQ}\APACrefatitle {Spatiotemporal Feedback and Network Structure Drive and
  Encode \textit{Caenorhabditis elegans} Locomotion} {Spatiotemporal feedback
  and network structure drive and encode \textit{Caenorhabditis elegans}
  locomotion}.{\BBCQ}
\newblock
\APACjournalVolNumPages{PLoS Comput Biol}{13}{1}{e1005303}.
\newblock
\begin{APACrefDOI} \doi{10.1371/journal.pcbi.1005303} \end{APACrefDOI}
\PrintBackRefs{\CurrentBib}

\bibitem [\protect \citeauthoryear {%
Kuramochi%
\ \BBA {} Doi%
}{%
Kuramochi%
\ \BBA {} Doi%
}{%
{\protect \APACyear {2017}}%
}]{%
Kuramochi2017}
\APACinsertmetastar {%
Kuramochi2017}%
\begin{APACrefauthors}%
Kuramochi, M.%
\BCBT {}\ \BBA {} Doi, M.%
\end{APACrefauthors}%
\unskip\
\newblock
\APACrefYearMonthDay{2017}{}{}.
\newblock
{\BBOQ}\APACrefatitle {A computational model based on multi-regional calcium
  imaging represents the spatio-temporal dynamics in a \textit{Caenorhabditis
  elegans} sensory neuron} {A computational model based on multi-regional
  calcium imaging represents the spatio-temporal dynamics in a
  \textit{Caenorhabditis elegans} sensory neuron}.{\BBCQ}
\newblock
\APACjournalVolNumPages{PLoS One}{12}{1}{e0168415}.
\newblock
\begin{APACrefDOI} \doi{10.1371/journal.pone.0168415} \end{APACrefDOI}
\PrintBackRefs{\CurrentBib}

\bibitem [\protect \citeauthoryear {%
Lebois%
\ \protect \BOthers {.}}{%
Lebois%
\ \protect \BOthers {.}}{%
{\protect \APACyear {2012}}%
}]{%
Lebois2012}
\APACinsertmetastar {%
Lebois2012}%
\begin{APACrefauthors}%
Lebois, F.%
, Sauvage, P.%
, Py, C.%
, Cardoso, O.%
, Ladoux, B.%
\BCBL {}\ \BBA {} Meglio, P\BPBI H\BPBI J.%
\end{APACrefauthors}%
\unskip\
\newblock
\APACrefYearMonthDay{2012}{}{}.
\newblock
{\BBOQ}\APACrefatitle {Locomotion control of \textit{Caenorhabditis elegans}
  through confinement} {Locomotion control of \textit{Caenorhabditis elegans}
  through confinement}.{\BBCQ}
\newblock
\APACjournalVolNumPages{Biophysical Journal}{102}{12}{2791--2798}.
\PrintBackRefs{\CurrentBib}

\bibitem [\protect \citeauthoryear {%
Lindsay%
, Thiele%
\BCBL {}\ \BBA {} Lockery%
}{%
Lindsay%
\ \protect \BOthers {.}}{%
{\protect \APACyear {2011}}%
}]{%
Lindsay2011}
\APACinsertmetastar {%
Lindsay2011}%
\begin{APACrefauthors}%
Lindsay, T\BPBI H.%
, Thiele, T\BPBI R.%
\BCBL {}\ \BBA {} Lockery, S\BPBI R.%
\end{APACrefauthors}%
\unskip\
\newblock
\APACrefYearMonthDay{2011}{}{}.
\newblock
{\BBOQ}\APACrefatitle {Optogenetic analysis of synaptic transmission in the
  central nervous system of the nematode \textit{Caenorhabditis elegans}}
  {Optogenetic analysis of synaptic transmission in the central nervous system
  of the nematode \textit{Caenorhabditis elegans}}.{\BBCQ}
\newblock
\APACjournalVolNumPages{Nat Commun}{2}{306}{}.
\newblock
\begin{APACrefDOI} \doi{10.1038/ncomms1304} \end{APACrefDOI}
\PrintBackRefs{\CurrentBib}

\bibitem [\protect \citeauthoryear {%
Lockery%
\ \BBA {} Goodman%
}{%
Lockery%
\ \BBA {} Goodman%
}{%
{\protect \APACyear {2009}}%
}]{%
Lockery2009}
\APACinsertmetastar {%
Lockery2009}%
\begin{APACrefauthors}%
Lockery, S\BPBI R.%
\BCBT {}\ \BBA {} Goodman, M\BPBI B.%
\end{APACrefauthors}%
\unskip\
\newblock
\APACrefYearMonthDay{2009}{}{}.
\newblock
{\BBOQ}\APACrefatitle {The quest for action potentials in \textit{C. elegans}
  neurons hits a plateau} {The quest for action potentials in \textit{C.
  elegans} neurons hits a plateau}.{\BBCQ}
\newblock
\APACjournalVolNumPages{Nat Neurosci}{12}{4}{377--378}.
\PrintBackRefs{\CurrentBib}

\bibitem [\protect \citeauthoryear {%
Majmudar%
, Keaveny%
, Zhang%
\BCBL {}\ \BBA {} Shelley%
}{%
Majmudar%
\ \protect \BOthers {.}}{%
{\protect \APACyear {2012}}%
}]{%
Majmudar2012}
\APACinsertmetastar {%
Majmudar2012}%
\begin{APACrefauthors}%
Majmudar, T.%
, Keaveny, E.%
, Zhang, J.%
\BCBL {}\ \BBA {} Shelley, M.%
\end{APACrefauthors}%
\unskip\
\newblock
\APACrefYearMonthDay{2012}{}{}.
\newblock
{\BBOQ}\APACrefatitle {Experiments and theory of undulatory locomotion in a
  simple structured medium} {Experiments and theory of undulatory locomotion in
  a simple structured medium}.{\BBCQ}
\newblock
\APACjournalVolNumPages{J R Soc Interface}{9}{73}{1809--1823}.
\PrintBackRefs{\CurrentBib}

\bibitem [\protect \citeauthoryear {%
Marder%
, Bucher%
, Schulz%
\BCBL {}\ \BBA {} Taylor%
}{%
Marder%
\ \protect \BOthers {.}}{%
{\protect \APACyear {2005}}%
}]{%
Marder2005}
\APACinsertmetastar {%
Marder2005}%
\begin{APACrefauthors}%
Marder, E.%
, Bucher, D.%
, Schulz, D\BPBI J.%
\BCBL {}\ \BBA {} Taylor, A\BPBI L.%
\end{APACrefauthors}%
\unskip\
\newblock
\APACrefYearMonthDay{2005}{}{}.
\newblock
{\BBOQ}\APACrefatitle {Invertebrate central pattern generation moves along}
  {Invertebrate central pattern generation moves along}.{\BBCQ}
\newblock
\APACjournalVolNumPages{Curr Biol}{15}{17}{R685--R699}.
\PrintBackRefs{\CurrentBib}

\bibitem [\protect \citeauthoryear {%
Mellem%
, Brockie%
, Madsen%
\BCBL {}\ \BBA {} Maricq%
}{%
Mellem%
\ \protect \BOthers {.}}{%
{\protect \APACyear {2008}}%
}]{%
Mellem2008}
\APACinsertmetastar {%
Mellem2008}%
\begin{APACrefauthors}%
Mellem, J\BPBI E.%
, Brockie, P\BPBI J.%
, Madsen, D\BPBI M.%
\BCBL {}\ \BBA {} Maricq, A\BPBI V.%
\end{APACrefauthors}%
\unskip\
\newblock
\APACrefYearMonthDay{2008}{}{}.
\newblock
{\BBOQ}\APACrefatitle {Action potentials contribute to neuronal signaling in
  \textit{C. elegans}} {Action potentials contribute to neuronal signaling in
  \textit{C. elegans}}.{\BBCQ}
\newblock
\APACjournalVolNumPages{Nat Neurosci}{11}{8}{865--867}.
\PrintBackRefs{\CurrentBib}

\bibitem [\protect \citeauthoryear {%
Niebur%
\ \BBA {} Erd\"{o}s%
}{%
Niebur%
\ \BBA {} Erd\"{o}s%
}{%
{\protect \APACyear {1991}}%
}]{%
Niebur1991}
\APACinsertmetastar {%
Niebur1991}%
\begin{APACrefauthors}%
Niebur, E.%
\BCBT {}\ \BBA {} Erd\"{o}s, P.%
\end{APACrefauthors}%
\unskip\
\newblock
\APACrefYearMonthDay{1991}{}{}.
\newblock
{\BBOQ}\APACrefatitle {Theory of the locomotion of nematodes: Dynamics of
  undulatory progression on a surface} {Theory of the locomotion of nematodes:
  Dynamics of undulatory progression on a surface}.{\BBCQ}
\newblock
\APACjournalVolNumPages{Biophys J}{60}{5}{1132--1146}.
\PrintBackRefs{\CurrentBib}

\bibitem [\protect \citeauthoryear {%
Niebur%
\ \BBA {} Erd\"{o}s%
}{%
Niebur%
\ \BBA {} Erd\"{o}s%
}{%
{\protect \APACyear {1993}}%
}]{%
Niebur1993}
\APACinsertmetastar {%
Niebur1993}%
\begin{APACrefauthors}%
Niebur, E.%
\BCBT {}\ \BBA {} Erd\"{o}s, P.%
\end{APACrefauthors}%
\unskip\
\newblock
\APACrefYearMonthDay{1993}{}{}.
\newblock
{\BBOQ}\APACrefatitle {Theory of the locomotion of nematodes: Control of the
  somatic motor neurons by interneurons} {Theory of the locomotion of
  nematodes: Control of the somatic motor neurons by interneurons}.{\BBCQ}
\newblock
\APACjournalVolNumPages{Math. Biosci}{118}{1}{51--82}.
\PrintBackRefs{\CurrentBib}

\bibitem [\protect \citeauthoryear {%
Qi%
, Garren%
, Shu%
, Tsien%
\BCBL {}\ \BBA {} Jin%
}{%
Qi%
\ \protect \BOthers {.}}{%
{\protect \APACyear {2012}}%
}]{%
qi2012photo}
\APACinsertmetastar {%
qi2012photo}%
\begin{APACrefauthors}%
Qi, Y\BPBI B.%
, Garren, E\BPBI J.%
, Shu, X.%
, Tsien, R\BPBI Y.%
\BCBL {}\ \BBA {} Jin, Y.%
\end{APACrefauthors}%
\unskip\
\newblock
\APACrefYearMonthDay{2012}{}{}.
\newblock
{\BBOQ}\APACrefatitle {Photo-inducible cell ablation in Caenorhabditis elegans
  using the genetically encoded singlet oxygen generating protein miniSOG}
  {Photo-inducible cell ablation in caenorhabditis elegans using the
  genetically encoded singlet oxygen generating protein minisog}.{\BBCQ}
\newblock
\APACjournalVolNumPages{Proceedings of the National Academy of
  Sciences}{109}{19}{7499--7504}.
\PrintBackRefs{\CurrentBib}

\bibitem [\protect \citeauthoryear {%
Rakowski%
, Srinivasan%
, Sternberg%
\BCBL {}\ \BBA {} Karbowski%
}{%
Rakowski%
\ \protect \BOthers {.}}{%
{\protect \APACyear {2013}}%
}]{%
Rakowski2013}
\APACinsertmetastar {%
Rakowski2013}%
\begin{APACrefauthors}%
Rakowski, F.%
, Srinivasan, J.%
, Sternberg, P.%
\BCBL {}\ \BBA {} Karbowski, J.%
\end{APACrefauthors}%
\unskip\
\newblock
\APACrefYearMonthDay{2013}{}{}.
\newblock
{\BBOQ}\APACrefatitle {Synaptic polarity of the interneuron circuit controlling
  \textit{C. elegans} locomotion} {Synaptic polarity of the interneuron circuit
  controlling \textit{C. elegans} locomotion}.{\BBCQ}
\newblock
\APACjournalVolNumPages{Front Comput Neurosci}{7}{}{128}.
\newblock
\begin{APACrefDOI} \doi{10.3389/fncom.2013.00128} \end{APACrefDOI}
\PrintBackRefs{\CurrentBib}

\bibitem [\protect \citeauthoryear {%
Sakata%
\ \BBA {} Shingai%
}{%
Sakata%
\ \BBA {} Shingai%
}{%
{\protect \APACyear {2004}}%
}]{%
Sakata2005}
\APACinsertmetastar {%
Sakata2005}%
\begin{APACrefauthors}%
Sakata, K.%
\BCBT {}\ \BBA {} Shingai, R.%
\end{APACrefauthors}%
\unskip\
\newblock
\APACrefYearMonthDay{2004}{}{}.
\newblock
{\BBOQ}\APACrefatitle {Neural network model to generate head swing in
  locomotion of \textit{Caenorhabditis elegans}} {Neural network model to
  generate head swing in locomotion of \textit{Caenorhabditis elegans}}.{\BBCQ}
\newblock
\APACjournalVolNumPages{Network}{15}{3}{199--216}.
\PrintBackRefs{\CurrentBib}

\bibitem [\protect \citeauthoryear {%
Selverston%
}{%
Selverston%
}{%
{\protect \APACyear {2010}}%
}]{%
Selverston2010}
\APACinsertmetastar {%
Selverston2010}%
\begin{APACrefauthors}%
Selverston, A\BPBI I.%
\end{APACrefauthors}%
\unskip\
\newblock
\APACrefYearMonthDay{2010}{}{}.
\newblock
{\BBOQ}\APACrefatitle {Invertebrate central pattern generator circuits}
  {Invertebrate central pattern generator circuits}.{\BBCQ}
\newblock
\APACjournalVolNumPages{Philos Trans R Soc Lond B Biol
  Sci}{365}{1551}{2329--2345}.
\PrintBackRefs{\CurrentBib}

\bibitem [\protect \citeauthoryear {%
Sulston%
\ \BBA {} Horvitz%
}{%
Sulston%
\ \BBA {} Horvitz%
}{%
{\protect \APACyear {1977}}%
}]{%
Sulston1977}
\APACinsertmetastar {%
Sulston1977}%
\begin{APACrefauthors}%
Sulston, J.%
\BCBT {}\ \BBA {} Horvitz, H.%
\end{APACrefauthors}%
\unskip\
\newblock
\APACrefYearMonthDay{1977}{}{}.
\newblock
{\BBOQ}\APACrefatitle {Post-embryonic cell lineages of the nematode,
  \textit{Caenorhabditis elegans}} {Post-embryonic cell lineages of the
  nematode, \textit{Caenorhabditis elegans}}.{\BBCQ}
\newblock
\APACjournalVolNumPages{Dev Biol}{56}{1}{110--156}.
\PrintBackRefs{\CurrentBib}

\bibitem [\protect \citeauthoryear {%
Varshney%
, Chen%
, Paniagua%
, Hall%
\BCBL {}\ \BBA {} Chklovskii%
}{%
Varshney%
\ \protect \BOthers {.}}{%
{\protect \APACyear {2011}}%
}]{%
Varshney2011}
\APACinsertmetastar {%
Varshney2011}%
\begin{APACrefauthors}%
Varshney, L.%
, Chen, B.%
, Paniagua, E.%
, Hall, D.%
\BCBL {}\ \BBA {} Chklovskii, D.%
\end{APACrefauthors}%
\unskip\
\newblock
\APACrefYearMonthDay{2011}{}{}.
\newblock
{\BBOQ}\APACrefatitle {Structural properties of the \textit{Caenorhabditis
  elegans} neuronal network} {Structural properties of the
  \textit{Caenorhabditis elegans} neuronal network}.{\BBCQ}
\newblock
\APACjournalVolNumPages{PLoS Comput Biol}{7}{2}{e1001066}.
\newblock
\begin{APACrefDOI} \doi{10.1371/journal.pcbi.1001066} \end{APACrefDOI}
\PrintBackRefs{\CurrentBib}

\bibitem [\protect \citeauthoryear {%
Waterston%
}{%
Waterston%
}{%
{\protect \APACyear {1988}}%
}]{%
Waterston1988}
\APACinsertmetastar {%
Waterston1988}%
\begin{APACrefauthors}%
Waterston, R.%
\end{APACrefauthors}%
\unskip\
\newblock
\APACrefYearMonthDay{1988}{}{}.
\newblock
{\BBOQ}\APACrefatitle {Muscle} {Muscle}.{\BBCQ}
\newblock
\BIn{} W.~Wood\ (\BED), \APACrefbtitle {The nematode \textit{C. elegans}} {The
  nematode \textit{C. elegans}}\ (\BPGS\ 281--335).
\newblock
\APACaddressPublisher{New York}{Cold Spring Harbor Laboratory Press}.
\PrintBackRefs{\CurrentBib}

\bibitem [\protect \citeauthoryear {%
Wen%
\ \protect \BOthers {.}}{%
Wen%
\ \protect \BOthers {.}}{%
{\protect \APACyear {2012}}%
}]{%
Wen2012}
\APACinsertmetastar {%
Wen2012}%
\begin{APACrefauthors}%
Wen, Q.%
, Po, M\BPBI D.%
, Hulme, E.%
, Chen, S.%
, Liu, X.%
, Kwok, S\BPBI W.%
\BDBL {}others%
\end{APACrefauthors}%
\unskip\
\newblock
\APACrefYearMonthDay{2012}{}{}.
\newblock
{\BBOQ}\APACrefatitle {Proprioceptive coupling within motor neurons drives
  \textit{C. elegans} forward locomotion} {Proprioceptive coupling within motor
  neurons drives \textit{C. elegans} forward locomotion}.{\BBCQ}
\newblock
\APACjournalVolNumPages{Neuron}{76}{4}{750--761}.
\PrintBackRefs{\CurrentBib}

\bibitem [\protect \citeauthoryear {%
White%
, Southgate%
, Thomson%
\BCBL {}\ \BBA {} Brenner%
}{%
White%
\ \protect \BOthers {.}}{%
{\protect \APACyear {1976}}%
}]{%
White1976}
\APACinsertmetastar {%
White1976}%
\begin{APACrefauthors}%
White, J.%
, Southgate, E.%
, Thomson, J.%
\BCBL {}\ \BBA {} Brenner, S.%
\end{APACrefauthors}%
\unskip\
\newblock
\APACrefYearMonthDay{1976}{}{}.
\newblock
{\BBOQ}\APACrefatitle {The structure of the ventral nerve cord of
  \textit{Caenorhabditis elegans}} {The structure of the ventral nerve cord of
  \textit{Caenorhabditis elegans}}.{\BBCQ}
\newblock
\APACjournalVolNumPages{Philos Trans R Soc Lond B Biol
  Sci}{275}{938}{327--348}.
\PrintBackRefs{\CurrentBib}

\bibitem [\protect \citeauthoryear {%
White%
, Southgate%
, Thomson%
\BCBL {}\ \BBA {} Brenner%
}{%
White%
\ \protect \BOthers {.}}{%
{\protect \APACyear {1986}}%
}]{%
White1986}
\APACinsertmetastar {%
White1986}%
\begin{APACrefauthors}%
White, J.%
, Southgate, E.%
, Thomson, J.%
\BCBL {}\ \BBA {} Brenner, S.%
\end{APACrefauthors}%
\unskip\
\newblock
\APACrefYearMonthDay{1986}{}{}.
\newblock
{\BBOQ}\APACrefatitle {The structure of the nervous system of the nematode
  \textit{Caenorhabditis elegans}} {The structure of the nervous system of the
  nematode \textit{Caenorhabditis elegans}}.{\BBCQ}
\newblock
\APACjournalVolNumPages{Philos Trans R Soc Lond B Biol
  Sci}{275}{938}{327--348}.
\PrintBackRefs{\CurrentBib}

\bibitem [\protect \citeauthoryear {%
Wicks%
, Roehrig%
\BCBL {}\ \BBA {} Rankin%
}{%
Wicks%
\ \protect \BOthers {.}}{%
{\protect \APACyear {1996}}%
}]{%
Wicks1996}
\APACinsertmetastar {%
Wicks1996}%
\begin{APACrefauthors}%
Wicks, S\BPBI R.%
, Roehrig, C\BPBI J.%
\BCBL {}\ \BBA {} Rankin, C\BPBI H.%
\end{APACrefauthors}%
\unskip\
\newblock
\APACrefYearMonthDay{1996}{}{}.
\newblock
{\BBOQ}\APACrefatitle {A dynamic network simulation of the nematode tap
  withdrawal circuit: predictions concerning synaptic function using behavioral
  criteria} {A dynamic network simulation of the nematode tap withdrawal
  circuit: predictions concerning synaptic function using behavioral
  criteria}.{\BBCQ}
\newblock
\APACjournalVolNumPages{J Neurosci}{16}{12}{4017--4031}.
\PrintBackRefs{\CurrentBib}

\bibitem [\protect \citeauthoryear {%
M.~Xu%
\ \protect \BOthers {.}}{%
M.~Xu%
\ \protect \BOthers {.}}{%
{\protect \APACyear {2013}}%
}]{%
Xu2013}
\APACinsertmetastar {%
Xu2013}%
\begin{APACrefauthors}%
Xu, M.%
, Jarrell, T\BPBI A.%
, Wang, Y.%
, Cook, S\BPBI J.%
, Hall, D\BPBI H.%
\BCBL {}\ \BBA {} Emmons, S\BPBI W.%
\end{APACrefauthors}%
\unskip\
\newblock
\APACrefYearMonthDay{2013}{}{}.
\newblock
{\BBOQ}\APACrefatitle {Computer assisted assembly of connectomes from electron
  micrographs: application to \textit{Caenorhabditis elegans}} {Computer
  assisted assembly of connectomes from electron micrographs: application to
  \textit{Caenorhabditis elegans}}.{\BBCQ}
\newblock
\APACjournalVolNumPages{PLoS One}{8}{1}{e54050}.
\newblock
\begin{APACrefDOI} \doi{10.1371/journal.pone.0054050} \end{APACrefDOI}
\PrintBackRefs{\CurrentBib}

\bibitem [\protect \citeauthoryear {%
T.~Xu%
\ \protect \BOthers {.}}{%
T.~Xu%
\ \protect \BOthers {.}}{%
{\protect \APACyear {2017}}%
}]{%
Xu2017}
\APACinsertmetastar {%
Xu2017}%
\begin{APACrefauthors}%
Xu, T.%
, Huo, J.%
, Shao, S.%
, Po, M.%
, Kawano, T.%
, Lu, Y.%
\BDBL {}Wen, Q.%
\end{APACrefauthors}%
\unskip\
\newblock
\APACrefYearMonthDay{2017}{}{}.
\newblock
{\BBOQ}\APACrefatitle {A descending pathway facilitates undulatory wave
  propagation in \textit{Caenorhabditis elegans} through gap junctions} {A
  descending pathway facilitates undulatory wave propagation in
  \textit{Caenorhabditis elegans} through gap junctions}.{\BBCQ}
\newblock
\APACjournalVolNumPages{bioRxiv}{}{}{}.
\newblock
\begin{APACrefDOI} \doi{10.1101/131490} \end{APACrefDOI}
\PrintBackRefs{\CurrentBib}

\bibitem [\protect \citeauthoryear {%
Zhen%
\ \BBA {} Samuel%
}{%
Zhen%
\ \BBA {} Samuel%
}{%
{\protect \APACyear {2015}}%
}]{%
Zhen2015}
\APACinsertmetastar {%
Zhen2015}%
\begin{APACrefauthors}%
Zhen, M.%
\BCBT {}\ \BBA {} Samuel, A\BPBI D.%
\end{APACrefauthors}%
\unskip\
\newblock
\APACrefYearMonthDay{2015}{}{}.
\newblock
{\BBOQ}\APACrefatitle {\textit{C. elegans} locomotion: small circuits, complex
  functions} {\textit{C. elegans} locomotion: small circuits, complex
  functions}.{\BBCQ}
\newblock
\APACjournalVolNumPages{Curr Opin Neurobiol}{33}{}{117--126}.
\PrintBackRefs{\CurrentBib}

\bibitem [\protect \citeauthoryear {%
Zheng%
, Brockie%
, Mellem%
, Madsen%
\BCBL {}\ \BBA {} Maricq%
}{%
Zheng%
\ \protect \BOthers {.}}{%
{\protect \APACyear {1999}}%
}]{%
Zheng1999}
\APACinsertmetastar {%
Zheng1999}%
\begin{APACrefauthors}%
Zheng, Y.%
, Brockie, P\BPBI J.%
, Mellem, J\BPBI E.%
, Madsen, D\BPBI M.%
\BCBL {}\ \BBA {} Maricq, A\BPBI V.%
\end{APACrefauthors}%
\unskip\
\newblock
\APACrefYearMonthDay{1999}{}{}.
\newblock
{\BBOQ}\APACrefatitle {Neuronal control of locomotion in \textit{C. elegans} is
  modified by a dominant mutation in the GLR-1 ionotropic glutamate receptor}
  {Neuronal control of locomotion in \textit{C. elegans} is modified by a
  dominant mutation in the glr-1 ionotropic glutamate receptor}.{\BBCQ}
\newblock
\APACjournalVolNumPages{Neuron}{24}{2}{347--361}.
\PrintBackRefs{\CurrentBib}

\end{thebibliography}

\end{document}